\documentclass[10pt,journal,a4paper,final,oneside,twocolumn]{IEEEtran}

\usepackage{times} 
\usepackage{graphicx}
\usepackage{cite}
\usepackage{citesort}
\usepackage{color}
\usepackage{psfrag}
\usepackage{makecell}
\usepackage{subfig}
\usepackage{caption}
\usepackage{amssymb}
\usepackage{stfloats}
\usepackage{lmodern}
\usepackage{epsfig}
\usepackage{pifont}
\usepackage{amsmath}

\usepackage{cases}
\usepackage{array}
\usepackage{multicol}
\usepackage{multirow}
\usepackage[T1]{fontenc}
\usepackage{comment}
\usepackage{enumerate}
\usepackage{amsthm}
\usepackage{indentfirst}
\usepackage{setspace}
\usepackage{bm}
\usepackage{float}
\usepackage{cases}
\usepackage[ruled, lined, linesnumbered, commentsnumbered, longend]{algorithm2e}
\SetKwInOut{KwIn}{Input}
\SetKwInOut{KwOut}{Output}
\SetKwProg{Init}{[Initialization]}{}{end}
\SetKwProg{QT}{[Applying the ELA method]}{}{end}
\SetKwProg{DT}{[Finding initial iteration values of UAV position and power coefficient]}{}{end}
\SetKwProg{LT}{[UAV trajectory and resource allocation optimization]}{}{end}
\SetKwProg{QB}{[Iteratively implementing the optimization at each RF]}{}{end}
\graphicspath{{figures/}}
\theoremstyle{remark}
\newtheorem{remark}{Remark}
\makeatletter
\renewcommand{\maketag@@@}[1]{\hbox{\m@th\normalsize\normalfont#1}}
\makeatother
\renewenvironment{thebibliography}[1]{
  \begin{oldthebibliography}{#1}
   \setlength{\parskip}{-0.1em}
}
{
  \end{oldthebibliography}
}

\makeatletter
\def\ifundefined{\@ifundefined}
\makeatother 

\begin{document}

\title{UAV-Enabled Integrated Sensing and Communication in Maritime Emergency Networks}

\author{Bohan Li,~\textit{Member, IEEE}, Jiahao Liu, Junsheng Mu, Pei Xiao,~\textit{Senior Member, IEEE}, Sheng Chen,~\textit{Life Fellow, IEEE} %
%\thanks{This work was supported in part by the National Key Research and Development Program of China under Grant 2024YFC3109100, and in part by the National Natural Science Foundation of China under Grant 62401530. ({\em Corresponding author: Jiahao Liu.})}
\thanks{B. Li and J. Liu are with the Faculty of Information Science and Engineering, Ocean University of China, Qingdao 266100, China (E-mails: bohan.li,ljh4327@ouc.edu.cn).}
\thanks{J. Mu is with the School of Information
and Communication Engineering, Beijing University of Posts and
Telecommunications, Beijing 100876, China (E-mails: mu\_junsheng@126.com)} 
\thanks{P. Xiao is with 5GIC \& 6GIC, University of Surrey, Guildford GU2 7XH, UK (E-mail: p.xiao@surrey.ac.uk)}%
\thanks{S. Chen is with the School of Electronics and Computer Science, University of Southampton, Southampton SO17 1BJ, UK (E-mail: sqc@ecs.soton.ac.uk).} %
\vspace*{-5mm}
}

\maketitle

\begin{abstract}
With line-of-sight mode deployment and fast response, unmanned aerial vehicle (UAV), equipped with the cutting-edge integrated sensing and communication (ISAC) technique, is poised to deliver high-quality communication and sensing services in maritime emergency scenarios. In practice, however, the real-time transmission of ISAC signals at the UAV side cannot be realized unless the reliable wireless fronthaul link between the terrestrial base station and UAV are available. This paper proposes a multicarrier-division duplex based joint fronthaul-access scheme, where mutually orthogonal subcarrier sets are leveraged to simultaneously support four types of fronthaul/access transmissions. In order to maximize the end-to-end communication rate while maintaining an adequate sensing quality-of-service (QoS) in such a complex scheme, the UAV trajectory, subcarrier assignment and power allocation are jointly optimized. The overall optimization process is designed in two stages. As the emergency area is usually far away from the coast, the optimal initial operating position for the UAV is first found. Once the UAV passes the initial operating position, the UAV's trajectory and resource allocation are optimized during the mission period to maximize the end-to-end communication rate under the constraint of minimum sensing QoS. Simulation results demonstrate the effectiveness of the proposed scheme in dealing with the joint fronthaul-access optimization problem in maritime ISAC networks, offering the advantages over benchmark schemes.
\end{abstract}


\vspace*{-2mm}
\section{Introduction}\label{S1}

Maritime rescue is never an easy task. Apart from natural factors, the limited capability of maritime communication and sensing (CAS) is the main challenge rendering the maritime rescue extremely difficult \cite{alqurashi2022maritime}. Without sufficiently high data rate communications and accurate sensing, the fast and effective rescue carried out several kilometers away from the coast is infeasible. To this end, advanced maritime emergency networks (MENs) have been investigated. In particular, the Global Maritime Distress and Safety System (GMDSS) is the most widely used maritime rescue system at present, which is composed of terrestrial base-station (TBS) and satellite networks, to provide long-range coverage of maritime communication and positioning services \cite{ilcev2020new}. However, subject to the long distance from TBS and satellite to emergency area, the GMDSS suffers from restricted data rate, low-resolution sensing and large latency, failing to meet the demand of wide-band communication and accurate sensing on the ocean.

By contrast, due to high flexibility and easy deployment, unmanned aerial vehicles (UAVs) are more competitive for constituting temporary emergency networks \cite{zhao2019uav}, and they can achieve higher transmission rate and enhanced communication coverage in the target area \cite{gu2023survey}. As for sensing, satellites and TBSs are better at searching the widely-unknown area, while UAVs are capable of quickly approaching the targets and implementing the high-resolution positioning and detection works once the location of the interested area is roughly known. Despite of these advantages, there is a paucity of works that leverage UAVs to simultaneously execute the emergency tasks of CAS on the ocean. This is mainly because at present concurrently employing both communication and radar equipments may impose significant energy burden on the UAV, leading to reduced mission period. However, it is not the case in the near future of Beyond 5G and 6G eras, as the integrated sensing and communication (ISAC) technology will enable the UAV to simultaneously carry out CAS on a common hardware platform, thereby largely saving energy consumption \cite{meng2023uav}. 

\vspace*{-4mm}
\subsection{Related Works}\label{S1.1}

\subsubsection{UAV-Aided MCNs}

To date, there have been many works studying the UAV-aided maritime communication networks (MCNs) \cite{nomikos2022survey}. The authors of \cite{lyu2021fast,qian2022joint,xin2023joint} proposed various UAV trajectory/deployment optimization and resource allocation algorithms to achieve different objectives in MCN, such as maximizing data collection capability of UAV \cite{lyu2021fast}, minimizing UAV's total energy consumption \cite{qian2022joint} and achieving the optimal system spectral efficiency \cite{xin2023joint}. However, the aforementioned works only consider the access link from the UAV to maritime users, but ignore the fronthaul link from the TBS to UAV. In fact, it is the fronthaul link that makes MCN different from terrestrial networks in the presence of UAV-aided communications. More specifically, in terrestrial networks, with the aid of handover technique among cells or sufficient bandwidth, e.g., millimeter-wave band, the UAV can usually connect to a BS in its close proximity or have plenty of spectral resource. Hence the performance of fronthaul link can be maintained \cite{zhai2023joint}. By contrast, for MCN, once the UAV flies far away from the TBS, its trajectory design and resource allocation must carefully consider the fronthaul link, as the degradation of which will highly hinder the achievable access rate. 
%Considering the importance of fronthaul links in MCN, the authors of \cite{li2020maritime} took the fronthaul constraint into account during UAV trajectory optimization. However, the proposed system applies the time-division strategy to implement fronthaul and access transmissions, which inevitably lowers the end-to-end data rate. 

\subsubsection{UAV-Enabled ISAC}

Driven by its features of on-demand deployment and line-of-sight (LoS)-dominant channels, the UAV-enabled ISAC (UAV-ISAC) has garnered tremendous attention recently \cite{meng2023uav}. The authors of \cite{liu2024trajectory,wang2020constrained,meng2022throughput,lyu2022joint,liu2023fair,deng2023beamforming} studied the optimization of UAV trajectory/deployment and resource allocation to maximize the communication data rate or minimize the energy consumption while attaining adequate quality-of-service (QoS) for sensing. The sensing-centered schemes have also been proposed in \cite{pan2023cooperative,zhang2022trajectory,liu2024uav,flores2024performance}, where the optimization aims to maximize the accuracy of localization or detection subject to adequate communication QoS. Instead of dealing with single-objective optimization, the authors of \cite{rezaei2023resource,jing2024isac,bayessa2024joint} concurrently optimized CAS relying on a weighted sum formulation. Here, three critical issues are worth further discussing. 

First, in UAV-ISAC related papers, the metrics of sensing performance can be generally classified into two categories, i.e., {information-theoretic metrics \cite{meng2022throughput,lyu2022joint,zhang2022trajectory,rezaei2023resource,liu2024trajectory,liu2024uav,zheng2024dual,bayessa2024joint,flores2024performance}, such as sensing mutual information (MI), sensing SINR and radar estimation rate,} and estimation-theoretic metrics \cite{wang2020constrained,deng2023beamforming,pan2023cooperative,jing2024isac}, such as Cram$\acute{\text{e}}$r-Rao bound  and mean square error. Information-theoretic metrics are independent of the estimator, which makes system optimization more general, while estimation-theoretic metrics are only used for explicitly characterizing the performance of specific sensing task. 

Second, in order to mitigate the interference between communication and sensing functions, most of the papers applied the time-division method to transmit communication signal and receive sensing echo at different time slots  {\cite{meng2022throughput,liu2023fair,deng2023beamforming,pan2023cooperative,liu2024trajectory,rezaei2023resource,zheng2024dual}, while the authors of \cite{zhang2022trajectory,flores2024performance} proposed the frequency-division method such that CAS signals can be transmitted over orthogonal bands.} Both these two methods suppress the interference at the expense of time or frequency resource. In addition, the reference \cite{lyu2022joint} implemented CAS using separate beams at the same time-frequency grid, but this approach inevitably causes residual digital-domain interference especially when targets and users are located at similar directions. 

{Third, only works  \cite{liu2024trajectory,wang2020constrained,zhang2022trajectory,liu2024uav,flores2024performance,zheng2024dual} considered the fronthaul links. In particular, \cite{liu2024trajectory,zhang2022trajectory,liu2024uav,zheng2024dual} leveraged the fronthaul links to feed the sensed information back to the data center, among which \cite{liu2024uav,zheng2024dual} used the access links to implement ISAC services, while \cite{liu2024trajectory,zhang2022trajectory} only transmitted sensing signal over the access links. On contrary, \cite{flores2024performance} assumed the sensing decision is only made at the UAV, and hence the fronthaul links are just employed for transmitting communication data. Note that, the above-mentioned papers adopted the decode-and-forward mode to implement sensing task, which requires the UAV to first process the sensed information based on its own baseband processor, and then forward the decoded sensing data to the data center. In this case, the UAV may suffer from extra computational overhead, resulting in the reduced flight time. Additionally, although the reference \cite{wang2020constrained} indeed considered a scenario where the UAV implements the ISAC transmission and exchanges the CAS information with BS, it neglected the effect of fronthaul links during the system optimization.}

%\begin{comment}
\begin{table*}[t]
\vspace*{-3mm}
\caption{Contrasting our proposed scheme with the literature of UAV-ISAC schemes}
\vspace{-2mm}
\scriptsize
\centering
{\begin{tabular}{|l|c|c|c|c|c|c|c|}
\hline
& {\bf Proposed} & \cite{pan2023cooperative} & \cite{zhang2022trajectory} & \cite{zheng2024dual}  & \cite{meng2022throughput,meng2023uav,liu2023fair,deng2023beamforming,rezaei2023resource} & \cite{lyu2022joint,jing2024isac,wang2020constrained,bayessa2024joint} & \cite{liu2024trajectory,flores2024performance,liu2024uav}  \\ \hline
Maritime scenario &  $\checkmark$ & &  & & &  &     \\ \hline
ISAC transmission & $\checkmark$ & $\checkmark$ & $\checkmark$ & $\checkmark$  & $\checkmark$ & $\checkmark$ & $\checkmark$         \\ \hline
End-to-end sensing MI &  $\checkmark$ & &  & & &  &   \\ \hline
Frame structure design & $\checkmark$& $\checkmark$  & & $\checkmark$ & $\checkmark$ & &    \\ \hline
FD-like waveform& $\checkmark$  &  & & & & &     \\ \hline
Joint fronthaul-access optimization & $\checkmark$ & & $\checkmark$& $\checkmark$ & & & $\checkmark$ \\ \hline
UAV's trajectory design &  $\checkmark$& $\checkmark$ & $\checkmark$ & $\checkmark$ & $\checkmark$& $\checkmark$ & $\checkmark$       \\ \hline
Joint power-subcarrier allocation & $\checkmark$& $\checkmark$& $\checkmark$  & &  &   &              \\ \hline
\end{tabular}}
\label{Table:MDDMar:compara} % Tab.I
\vspace*{-3mm}
\end{table*}
%\end{comment}

\vspace*{-4mm}
\subsection{Motivations and Contributions}\label{S1.2}
\subsubsection{Motivations}
Against the above background, in this paper, we exploit the UAV-ISAC technique in MENs, which has not been well-studied in open literature to the best of our knowledge. To make the application of UAV-ISAC in maritime emergency scenarios a reality, there are three challenges to be properly solved. {(i)~To satisfy the requirements of high transmission rate and low end-to-end latency in MENs, the full-duplex (FD) design of ISAC waveform is necessary. In this case, the interference between communication and sensing within the same time slot and frequency band has to be carefully addressed. (ii)~Maritime emergency usually happens far away from the coast, and hence the performance of the long-range wireless fronthaul links between the UAV and TBS is essential. In other words, to ensure the high-quality real-time downlink (DL) communication and target sensing, the fronthaul and access links must be jointly optimized in MENs. (iii)~As the timely CAS services are of paramount importance in MENs, it would be too late to provide services upon UAV arriving at the emergency spot. Instead, the feasible services should be offered as early as possible after UAV taking off. Therefore, in order to achieve timely CAS in MENs, the UAV's trajectory has to be specifically designed. These challenges motivate us to design a tailor-made ISAC waveform and frame structure for MENs, and jointly consider the fronthaul and access links during the optimization of resource allocation and UAV's trajectory, so as to obtain the optimal system performance.  }

\subsubsection{Contributions}
The novelties and contributions of this paper are summarized as follows.
\begin{itemize}
\item We design a UAV-ISAC scheme for MENs. To mitigate the interference between DL communication and sensing echo signals and improve the end-to-end performance, the multicarrier-division duplex (MDD)-based ISAC waveform is proposed, in which the TBS-to-UAV fronthaul link, UAV-to-TBS fronthaul link, UAV-to-user DL link and UAV-to-target sensing link are assigned with four mutually orthogonal subcarrier sets. To compensate for the frequency loss caused by subcarrier-division operation, an advanced multi-stream frame structure is tailor-made for the proposed networks, thanks to the FD characteristic of MDD \cite{li2021multicarrier}.
\item Considering the requirements of real-time CAS services in emergency area, the fronthaul links between the UAV and the TBS are practically modeled. Two different operating modes, decode-and-forward and amplify-and-forward, are applied at the UAV side to process the coded DL data from the TBS and the perceived information to the TBS, respectively. The end-to-end sensing MI between targets and the TBS based on amplify-and-forward mode is derived to evaluate the sensing performance, which has not been studied in the existing UAV-ISAC scenarios. 
\item In order to maximize the end-to-end DL rate while maintaining adequate sensing QoS, two sub-problems with respect to the optimization of UAV trajectory, power allocation at UAV and TBS sides, and subcarrier assignment within fronthaul and access links are presented. More specifically, considering the fact that the UAV cannot implement the CAS immediately after taking off due to the long distance away from the interested area, the first sub-problem aims to find the UAV's optimal initial operating location. Then, the second sub-problem maximizes the end-to-end DL rate under the constraint of minimum sensing MI during the mission period.
\end{itemize}
{Finally, a brief comparison between our proposed scheme with the existing UAV-ISAC works is shown in Table \ref{Table:MDDMar:compara}.}

\begin{figure}[t]
\centering
\includegraphics[width=0.6\linewidth]{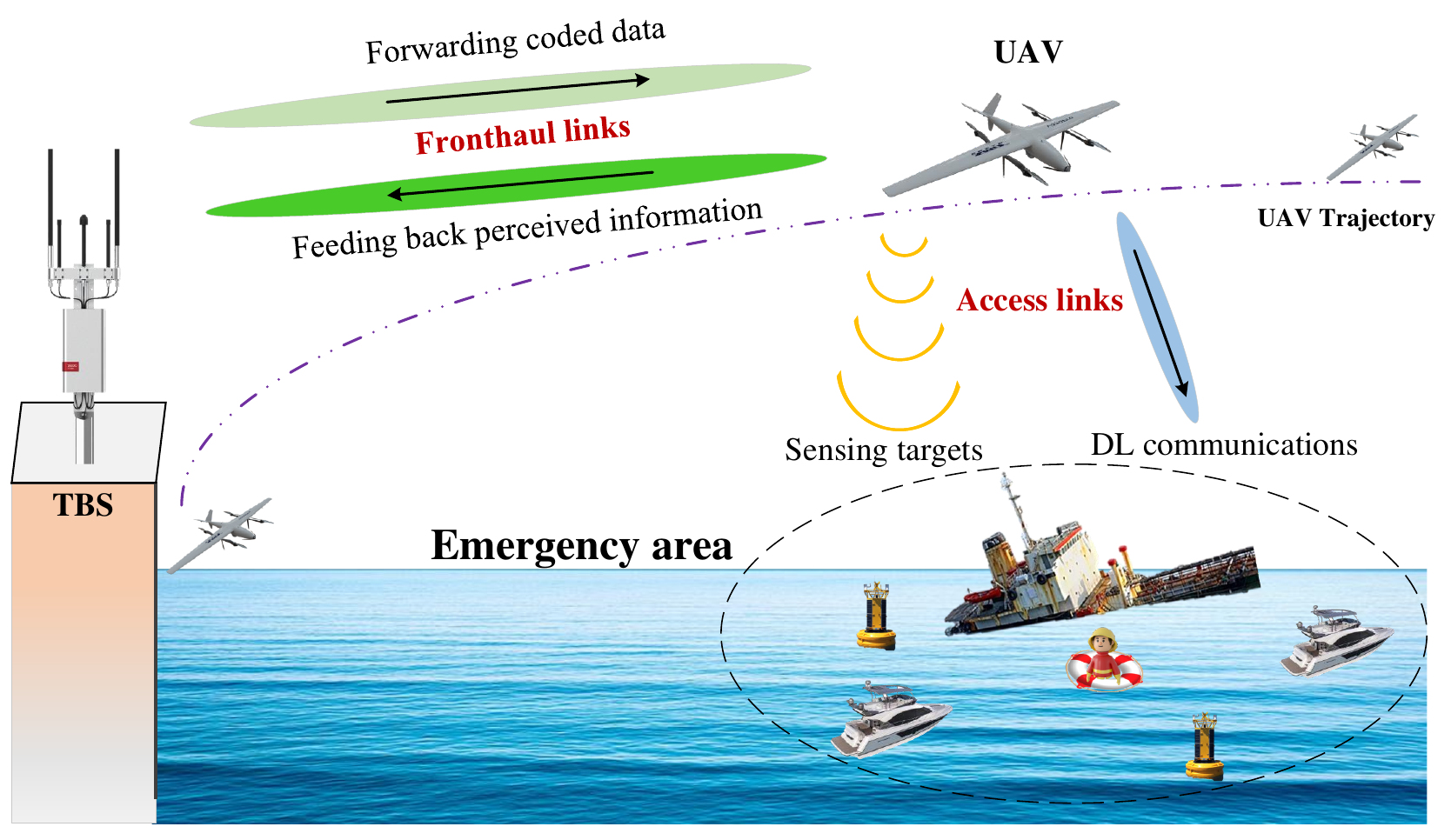}
\vspace*{-2mm}
\caption{\small UAV-ISAC enabled maritime emergency network.}
\label{figure-MDDMari-scen} % Fig.1
\vspace*{-3mm}
\end{figure}

\vspace*{-3mm}
\section{System Model}\label{S2}

Consider a UAV-assisted maritime emergency network which includes a TBS, a fixed-wing UAV, \scalebox{0.9}{$U$} mobile users (ships requiring communication service) constituting the set \scalebox{0.9}{$\mathcal{U}$}, and \scalebox{0.9}{$J$} maritime targets (buoys, ships and other surface vehicles) constituting the set \scalebox{0.9}{$\mathcal{J}$}, as shown in Fig.~\ref{figure-MDDMari-scen}. The on-demand UAV is deployed near the coast. Once the emergency occurs, the UAV flies toward the designated area to constitute an airborne network. Equipped with ISAC technique, the UAV concurrently communicates with mobiles users and sense targets at the same time and frequency band. The TBS is based along the coast and acts as the CPU establishing wireless fronthaul links with the UAV for forwarding the {source data} to and receiving the perceived information from the UAV. To meet the stringent energy limit of UAV and avoid the overhead of extra radio-frequency (RF) components at the UAV, the fronthaul and access links share the spectrum. 

Both the TBS and UAV operate in MDD mode. Specifically, all the subcarriers, defined by the index set \scalebox{0.9}{$\{m|m\! \in\!\mathcal{M}, |\mathcal{M}|\! =\!M\}$} within the frequency band, are coarsely classified into two blocks, i.e., \scalebox{0.9}{$\mathcal{M}^{\text{D}}$ and $\mathcal{M}^{\text{S}}$}, for implementing DL- and sensing-related tasks, respectively. Assumed that communication and sensing are of equal importance in the proposed maritime emergency scenario, the first $M/2$ subcarriers within \scalebox{0.9}{$\mathcal{M}$} constitute \scalebox{0.9}{$\mathcal{M}^{\text{D}}$}, while the last \scalebox{0.9}{$M/2$} subcarriers constitute \scalebox{0.9}{$\mathcal{M}^{\text{S}}$}. At the $n$-th radio frame, \scalebox{0.9}{$\mathcal{M}^{\text{D}}$} is further divided into \scalebox{0.9}{$\mathcal{M}_n^{\text{SD}}$} and \scalebox{0.9}{$\mathcal{M}_n^{\text{DL}}$}, which are used to transmit source data at the TBS and implement DL communications at the UAV, respectively. Similarly, \scalebox{0.9}{$\mathcal{M}^{\text{S}}$} is further cut into \scalebox{0.9}{$\mathcal{M}_n^{\text{SEN}}$} and \scalebox{0.9}{$\mathcal{M}_n^{\text{PE}}$} for the UAV to carry out sensing and feed back the perceived signal to the TBS, respectively. Note that the way of fine division of \scalebox{0.9}{$\mathcal{M}^{\text{D}}$} and \scalebox{0.9}{$\mathcal{M}^{\text{S}}$} is flexible and time-variant, dependent on the specific scenario. Denote by \scalebox{0.9}{$\alpha_{n,m}^{\text{X}}\! \in\! \left\{0,1\right\}$, $\text{X}\! \in\! \left\{\text{DL, SEN, PE, SD}\right\}$}, the indicator of subcarrier assignment at the $n$-th radio frame. If \scalebox{0.9}{$\alpha_{n,m}^{\text{X}}\! =\! 1$}, then \scalebox{0.9}{$m\! \in\! \mathcal{M}_n^{\text{X}}$}. According to the principle of MDD, we have\setcounter{equation}{0}\vspace{-1mm}

\begin{figure}[t]
\vspace*{-2mm}
\centering
\includegraphics[width=0.6\linewidth]{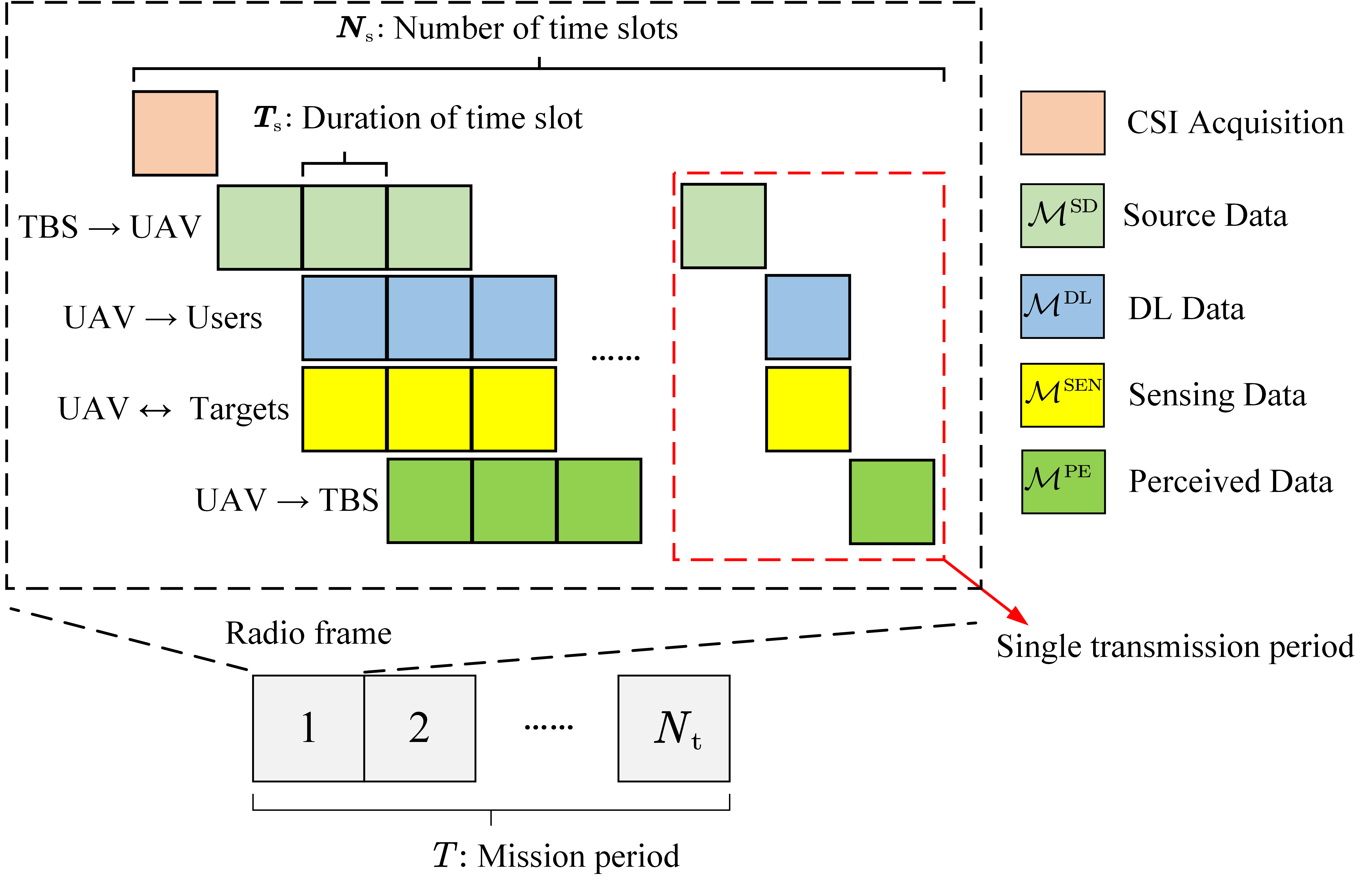}
\vspace*{-3mm}
\caption{\small Frame structure of UAV-enabled maritime emergency network operating on MDD mode.}
\label{figure-MDDMari-FS} % Fig.2
\end{figure}

{\begin{small}
\begin{equation}\label{eq:MDDMar:ISACSig} % eq.1
  \alpha_{n,m}^{\text{DL}}+\alpha_{n,m}^{\text{SEN}}+\alpha_{n,m}^{\text{PE}}+\alpha_{n,m}^{\text{SD}}\leq 1, \forall n, m\in\mathcal{M}.
\end{equation}
\end{small}}%
The frame structure is illustrated in Fig.~\ref{figure-MDDMari-FS}, where the mission period $T$ consists of $N_{\text{t}}$ radio frames, and each radio frame includes $N_{\text{s}}$ time slots. It is assumed that during each radio frame, the UAV and mobile users are quasi-stationary and stay at fixed locations, while the channel state information (CSI) of fronthaul and access channels remain unchanged. Hence the transmission procedure during a single transmission period can be described as follows. 

\emph{1)}~The UAV receives the pilots from users and derives the CSI of DL channels\footnote{As we mainly study the MDD-aided optimization trade-off among fronthaul and access links in maritime ISAC scenarios, the perfect CSI is assumed as in \cite{zheng2024dual,liu2024uav,meng2022throughput}, to avoid the deviation of the core of this paper.}. 

\emph{2)}~The UAV sends pilots and DL CSI to the TBS, who then estimates the CSI of the channel from the UAV to itself. Since the UAV and TBS leverage two orthogonal subcarrier sets within the same frequency band for two-way fronthaul transmissions, the two-way fronthaul channels exhibit time-domain reciprocity and frequency-domain correlation. Consequently, the TBS can derive the CSI of the channel from itself to the UAV. 

\emph{3)}~The TBS sends source data to the UAV based on the users' request, and feeds back the CSI of fronthaul channel to the UAV. In addition, as the TBS may have access to the rough image of emergency area via satellite remote sensing, it can provide the UAV with approximate coordinate of interested area so that UAV can implement quick sensing. Note that the delivery of CSI and auxiliary sensing coordinates can be achieved via control channels, and therefore will not affect system optimization. 

\emph{4)}~{The UAV adopts the decode-and-forward relay mode, which enables the UAV to firstly decode the data received from the TBS, and then forward it to users. Depending on the {\em{a-priori}} information of targets' locations, the UAV leverages DL data to carry out sensing tasks\footnote{Although the targets' locations are known at the current time slot, they may be changed at the next time slot. Hence, the sensing tasks aim at detecting whether the targets are still located at the same positions, or tracking the targets' new positions based on the previous ones.}.}

\emph{5)}~{The UAV resorts to the amplify-and-forward relay mode, by which the UAV is able to firstly amplify the echo signals, and then transmit the perceived information to the TBS for final sensing decision.}

\begin{figure*}[!t]\setcounter{equation}{1}
\footnotesize
\vspace*{-4mm}
\begin{equation}\label{eq:MDDMar:UAVTx} % eq.2
  \bm{s}[n,t,m] = \left\{ \begin{array}{cl}
    \bm{0}, & 1\leq t \leq 2, \\
    \alpha_{n,m}^{\text{SEN}} \sum\nolimits_{j=1}^J \bm{w}_{\text{SEN}}^j[n,m] x_{\text{SEN}}^j[n,t,m] + \alpha_{n,m}^{\text{DL}} \sum\nolimits_{u=1}^U \bm{w}_{\text{DL}}^u[n,m] x_{\text{DL}}^u[n,t,m], & t=3, \\
    \alpha_{n,m}^{\text{SEN}} \sum\nolimits_{j=1}^J \bm{w}_{\text{SEN}}^j[n,m] x_{\text{SEN}}^j[n,t,m]+\alpha_{n,m}^{\text{DL}}\sum\nolimits_{u=1}^U\pmb{w}_{\text{DL}}^u[n,m]x_{\text{DL}}^u[n,t,m] +\alpha_{n,m}^{\text{PE}} \sum\nolimits_{j=1}^J \bm{w}_{\text{PE}}^j[n,m] x _{\text{PE}}^j[n,t,m], & t\geq 4 .
  \end{array} \right.
\end{equation}  
\hrulefill
\vspace*{-1mm}
\begin{align}\label{eq:MDDMar:UAVRxSenE} % eq.3
\bm{r}_{\text{PE}}[n,t,m] = \big[x_{\text{PE}}^1[n,t,m],...,x_{\text{PE}}^J[n,t,m]\big]^{\text{T}}
%\left[\begin{matrix} x_{\text{PE}}^1[n,t,m] \\ \vdots \\
 %   x_{\text{PE}}^J[n,t,m] \\ \end{matrix} \right]
  =\left\{\begin{array}{cl}
    \bm{0}, & 1\leq t \leq 2, \\
    \alpha_{n,m}^{\text{SEN}} \widetilde{\bm{W}}_{\text{SEN}}^{\rm H}[n,m] \bm{H}_{\text{SEN}}^{\rm H}[n,m] \bm{\Lambda}_{\text{SEN}}[n,m] \bm{x}_{\text{SEN}}[n,t,m]  & \\
		+ \widetilde{\bm{W}}_{\text{SEN}}^{\rm H}[n,m] \left(\bm{n}_{\text{UAV}}[n,m] + \bm{z}_{\text{SI}}[n,m]\right), & t\geq 3 .
  \end{array} \right.
\end{align}
\hrulefill
\vspace*{-7mm}
\end{figure*}

\begin{figure*}[!b]\setcounter{equation}{4}
\footnotesize
\vspace*{-5mm}
\hrulefill
\vspace*{-1mm}
\begin{align}\label{eq:MDDMar:yDL} % eq.5
  y_u[n,t,m]\! =\! \left\{\!\!\! \begin{array}{cl}
    0, \!\! &\!\!\!\!\hspace*{-5mm} 1 \leq t \leq 2, \\
    \alpha_{n,m}^{\text{DL}}\left(\bm{h}_{\text{DL}}^{u}[n,m]\right)^{\rm H} \bm{w}_{\text{DL}}^u[n,m] x_{\text{DL}}^u[n,t,m] + \alpha_{n,m}^{\text{DL}}\!\!\!\! \sum\limits_{u^{\prime}=1,u^{\prime}\neq u}^U\!\!\!\! \left(\bm{h}_{\text{DL}}^{u}[n,m]\right)^{\rm H} \bm{w}_{\text{DL}}^{u^{\prime}}[n,m] x_{\text{DL}}^{u^{\prime}}[n,t,m] \!\! + n_u[n,m], \!\! &\!\! t\geq 3 ,
  \end{array}\right.\!\!
\end{align}
\vspace*{-4mm}
\end{figure*}

\vspace*{-2mm}
\subsection{Communication and Sensing Links}\label{2.1}

The UAV is equipped with \scalebox{0.9}{$R_{\text{tx}}\! =\! R_{\text{tx}}^{\text{l}}\! \times\! R_{\text{tx}}^{\text{w}}$} transmit and \scalebox{0.9}{$R_{\text{rx}}\! =\! R_{\text{rx}}^{\text{l}}\! \times\! R_{\text{rx}}^{\text{w}}$} receive uniform planar arrays (UPAs), which are placed parallel to the ground and sea surface. As shown in Fig.~\ref{figure-MDDMari-FS}, at the $t$-th time slot of the $n$-th radio frame, the integrated fronthaul, communication and sensing signal transmitted by the UAV over the $m$-th subcarrier can be expressed as \eqref{eq:MDDMar:UAVTx} at the top of the page, where \scalebox{0.9}{$\bm{w}_{\text{SEN}}^j[n,m]$ ($x_{\text{SEN}}^j[n,t,m]$), $\bm{w}_{\text{DL}}^u[n,m]$ ($x_{\text{DL}}^u[n,t,m]$)} and \scalebox{0.9}{$\bm{w}_{\text{PE}}^j[n,m]$ (${x}_{\text{PE}}^j[n,t,m]$)} denote the precoders (data) for sensing target $j$, communicating with user $u$ and sending the perceived information of target $j$ back to the TBS, respectively, with each data having unit energy, i.e., \scalebox{0.9}{$\mathbb{E}\left[|x[n,t,m]|^2\right]\! =\! 1$}. The transmit power of the UAV is constrained by \scalebox{0.9}{$\sum_{m=1}^{M}\mathbb{E}\left[\big\|\bm{s}[n,t,m]\right\|^2\big]\! \leq\! P_{\text{UAV}}$}.
{ 
\begin{remark}\label{Rk1}
Since the UAV receives the source data and sensing echo while transmitting integrated signal, it suffers from self-interference (SI). We assume that the UAV can rely on the combination of passive cancellation methods (e.g., implementing antenna separation between transmit and receive arrays \cite{debaillie2014analog}, and placing a radio frequency absorber among transceiver \cite{everett2014passive}) and active cancellation methods (e.g., multi-tap RF canceller \cite{kolodziej2016multitap} and adaptive beamforming-aided suppression \cite{li2020self}) to provide sufficient analog-domain SI cancellation (SIC) such that the power of SI falls into the dynamic range of ADC. Then, due to the characteristic of MDD \cite{li2021multicarrier}, the reception of source signal is free from the digital-domain SI with the aid of fast Fourier transform (FFT) operation. As for the reception of echo signal, although the digital interference components of DL and perceived signal can be readily removed by FFT, the transmitted sensing signal directly arriving at the receiver gives rise to the digital-domain SI, which can be modeled as \scalebox{0.9}{$\bm{z}_{\text{SI}}[n,m]\! \in\!\mathbb{C}^{R_{\text{rx}}}\! \sim\! \mathcal{CN}(\bm{0},\alpha_{n,m}^{\text{SEN}} \xi_{\text{SIC}} \sum_{j=1}^J\|\bm{w}_{\text{SEN}}^j[n,m]\|^2\pmb{I}_{R_{\text{rx}}})$}, where \scalebox{0.9}{$\xi_{\text{SIC}}$} denotes the SIC capability at the UAV taking account the analog- and digital-domain SI suppression.
\end{remark}}
At the UAV, the received echo signal reflected from target $j$ at the $m$-th subcarrier is given in \eqref{eq:MDDMar:UAVRxSenE} at the top of the page, where \scalebox{0.9}{$\bm{x}_{\text{SEN}}[n,t,m]\! =\! [x_{\text{SEN}}^1[n,t,m],\cdots ,x_{\text{SEN}}^J[n,t,m]]^{\rm T}$}, \scalebox{0.9}{$\bm{\Lambda}_{\text{SEN}}[n,m]=\! \text{diag}(\bm{W}_{\text{SEN}}[n,m])\!\in\! \mathbb{C}^{R_{\rm tx}J\times J}$} with \scalebox{0.9}{$\bm{W}_{\text{SEN}}[n,m]\!$} \scalebox{0.9}{$=\! [\bm{w}_{\text{SEN}}^1[n,m],\cdots,\bm{w}_{\text{SEN}}^J[n,m]]\! \in\! \mathbb{C}^{R_{\rm tx}\times J}$} and \scalebox{0.9}{$\rm{diag}(\cdot)$} denoting the block diagonal transformation, \scalebox{0.9}{$\widetilde{\bm{W}}_{\text{SEN}}[n,m]\!$} \scalebox{0.9}{$=\! \widetilde{\bm{w}}_{\text{SEN}}^1[n,m],\cdots ,\widetilde{\bm{w}}_{\text{SEN}}^J[n,m]]\! \in\! \mathbb{C}^{R_{\rm rx}\times J}$} with \scalebox{0.9}{$\widetilde{\bm{w}}_{\text{SEN}}^j[n,m]\! \in\! \mathbb{C}^{R_{\rm rx}}$} denoting the echo combining vector for target \scalebox{0.9}{$j$}, and \scalebox{0.9}{$\bm{H}_{\text{SEN}}[n,m]\! =\! [(\bm{H}_{\text{SEN}}^1[n,m])^{\rm H},\cdots ,(\bm{H}_{\text{SEN}}^J[n,m])^{\rm H}]^{\rm H}\!\in\! \mathbb{C}^{R_{\rm tx}J\times R_{\rm rx}}$} is the concentrated sensing channels of targets with \scalebox{0.9}{$\bm{H}_{\text{SEN}}^j[n,m]\! \in\! \mathbb{C}^{R_{\rm tx}\times R_{\rm rx}}$} given in \eqref{eqLoSlink}, while \scalebox{0.9}{$\bm{n}_{\text{UAV}}[n,m]\! \sim\! \mathcal{CN}\big(\bm{0},N_0\pmb{I}_{R_{\text{rx}}}\big)$} denotes the additive white Gaussian noise with \scalebox{0.9}{$N_0$} representing noise power spectrum density.
As we assume that targets are sparsely distributed, the terms of inter-beam interference caused by transmit antenna sidelobes, i.e., \scalebox{0.9}{$\big\{\widetilde{\bm{w}}_{\text{SEN}}^j[n,m]\big(\bm{H}_{\text{SEN}}^a[n,m]\big)^{\rm H} \bm{w}_{\text{SEN}}^b[n,m] x_{\text{SEN}}^b[n,t,m] \big| $} \scalebox{0.9}{$ \forall a, b\! \in\! \mathcal{J} \text{ and } a\! \neq\! b\big\}$}, are omitted to facilitate analysis. Unlike the complex scattering environment around TBS, the channels between the UAV and maritime users and targets are dominated by LoS link. Therefore, the sensing channel \scalebox{0.9}{$\bm{H}_{\text{SEN}}^{j}[n,m]$} is given by\setcounter{equation}{3} \vspace{-2mm}

\begin{small}
\begin{align}  \label{eqLoSlink} % eq.4
& \bm{H}_{\text{SEN}}^{j}[n,m] = \sqrt{\frac{G_{\mathrm{UAV}}^{\mathrm{tx}} G_{\mathrm{UAV}}^{\mathrm{rx}} \lambda_m^2 \sigma^{\mathrm{RCS}}_j}{( 4\pi )^3 d_{j}^{4}[n]}} \varpi_{j,n,m} e^{-\textsf{j} 2\pi \tau_j f_m} \times \nonumber \\
  & \, e^{\textsf{j} 2\pi f_{\mathrm{D},j} t_0} \bm{\alpha}_{\mathrm{tx}}\big( \theta_{j}^{\mathrm{tx}}[n],\phi_{j}^{\mathrm{tx}}[n],m\big) \bm{\alpha }_{\mathrm{rx}}^{\rm H}\big( \theta_{j}^{\mathrm{rx}}[n],\phi_{j}^{\mathrm{rx}}[n],m\big),\!
\end{align}
\end{small}%
where \scalebox{0.9}{$G_{\mathrm{UAV}}^{\mathrm{tx}}$, $G_{\mathrm{UAV}}^{\mathrm{rx}}$, $\lambda_m$, $\sigma^{\mathrm{RCS}}_j$, $\tau_j$, $t_0$, $\varpi_{j,n,m}\! \sim\! \mathcal{CN}\left(0,1\right)$, $f_m$} and \scalebox{0.9}{$f_{\text{D},j}$ }are the UAV transmitter and receiver antenna gains, wavelength, radar cross-section (RCS), path delay, symbol duration, complex gain, center frequency of the $m$-th sub-band and Doppler frequency, respectively. The Doppler and delay effects are assumed to be perfectly compensated through synchronization \cite{meng2022throughput}. The distance between the UAV and target $j$ is given by \scalebox{0.9}{$d_j[n]\! =\! \big\|\bm{c}_{\text{UAV}}[n]\! -\! \bm{c}_j\big\|$}, where \scalebox{0.9}{$\bm{c}_{\text{UAV}}[n]\! =\! \big[x_{\text{UAV}}[n],y_{\text{UAV}}[n],z_{\text{UAV}}\big]$} and \scalebox{0.9}{$\pmb{c}_j\! =\! \big[x_j,y_j,z_j\big]$} denote the coordinates of the UAV and the $j$-th target at the $n$-th radio frame, respectively. The fly height of the UAV is assumed to be constant. Moreover, \scalebox{0.9}{$\bm{\alpha}_{\mathrm{tx}}(\cdot )$} and \scalebox{0.9}{$\bm{\alpha}_{\mathrm{rx}}(\cdot )$} denote the UPA response vectors of the UAV transmitter and receiver, respectively.

During the $t$-th time slot of the $n$-th radio frame, the DL communication signal of subcarrier $m$ received at user $u$ is given in (\ref{eq:MDDMar:yDL}) at the bottom of the previous page, where \scalebox{0.9}{$n_u[n,m]\sim\mathcal{CN}(0,N_0)$} and the DL communication channel is modeled as\setcounter{equation}{5} \vspace{-3mm}

\begin{small}
\begin{align}\label{eqDLs} % eq.6
  \bm{h}_{\text{DL}}^u[n, m] =& \sqrt{\frac{G_{\mathrm{UAV}}^{\mathrm{tx}} G_{\mathrm{UE}}^{\mathrm{rx}} \lambda_m^2}{( 4\pi )^2 d_u^{2}[n]}} \varpi_{u,n,m} e^{-\textsf{j}2\pi \tau_u f_m} e^{\textsf{j}2\pi f_{\mathrm{D},u}t_0} \nonumber \\
  & \times \bm{\alpha}_{\mathrm{tx}}\big( \theta^{\mathrm{tx}}_u[n],\phi^{\mathrm{tx}}_u[n],m\big) \in \mathbb{C}^{R_{\text{tx}}},
\end{align}
\end{small}%
in which \scalebox{0.9}{$\varpi_{u,n,m}\! \sim\! \mathcal{CN}(0,1)$, $d_u[n]\! =\! \left\|\bm{c}_{\text{UAV}}[n]\! -\! \bm{c}_u\right\|_2$} with \scalebox{0.9}{$\bm{c}_u\! =\! [x_u,y_u,z_u]$} denoting the coordinate of user $u$. As the subcarriers used for transmitting the source, sensing and perceived signals are orthogonal to that used for DL transmissions, users only suffer from the co-subchannel multi-user interference.

\vspace*{-2mm}
\subsection{Two-Way Fronthaul Links}\label{2.2}

The TBS sends the source data to the UAV using \scalebox{0.9}{$\bar{R}_{\text{tx}}\! =\! \bar{R}_{\text{tx}}^{\text{l}}\! \times\! \bar{R}_{\text{tx}}^{\text{w}}$} transmit UPA. The UAV works as decode-and-forward relay to firstly decode the received data and then carry out CAS tasks. Due to the complex sensing environment and limited computation resource, the UAV adopts amplify-and-forward method in sensing task, that is, it only amplifies the received target echo and directly forwards it to the TBS for sensing data processing. To distinguish the perceived information of different targets, the TBS equips with \scalebox{0.9}{$\bar{R}_{\text{rx}}\! =\! \bar{R}_{\text{rx}}^{\text{l}}\! \times\! \bar{R}_{\text{rx}}^{\text{w}}$} receive UPA. 

\begin{figure*}[!b]\setcounter{equation}{6}
\footnotesize
\vspace*{-5mm}
\hrulefill
\vspace*{-2mm}
\begin{align}\label{eq:MDDMar:UAVRxDL} % eq.7
  & \bm{y}_{\text{SD}}[n,t,m] = \left\{ \begin{array}{cl}
    \bm{0}, & \hfill t=1, \\
    \alpha_{n,m}^{\text{SD}} \widetilde{\bm{W}}_{\text{SD}}^{\rm H}[n,m] \bm{H}_{\text{SD}}^{\rm H}[n,m] \bm{F}_{\text{SD}}[n,m] \bm{x}_{\text{SD}}[n,t,m] + \widetilde{\bm{W}}_{\text{SD}}^{\rm H}[n,m] \bm{n}_{\text{UAV}}[n,m] , & t\geq 2,
  \end{array}\right. \!\!
\end{align}
\hrulefill
\vspace*{-2mm}
\begin{align}\label{eq:MDDMar:hBU} % eq.8
  \bm{H}_{\mathrm{SD}}[n, m] =& \sqrt{\frac{G_{\mathrm{TBS}}^{\mathrm{tx}} G_{\mathrm{UAV}}^{\mathrm{rx}} \lambda_m ^2}{( 4\pi )^2d_{\mathrm{TU}}^{2}[n]}} e^{-\textsf{j}2\pi \tau f_m} e^{\textsf{j}2\pi f_{\mathrm{D}}t_0} \big( \sqrt{\frac{K_{\mathrm{TU}}}{K_{\mathrm{TU}}+1}} \varpi_{n,m} \bm{\alpha}_{\mathrm{tx}}^{\mathrm{TBS}} \left( \theta^{\mathrm{tx}}[n],\phi^{\mathrm{tx}}[n],m\right) \bm{\alpha}_{\mathrm{rx}}^{\rm H} \left( \theta^{\mathrm{rx}}[n],\phi^{\mathrm{rx}}[n],m\right)  + \sqrt{\frac{1}{K_{\mathrm{TU}}+1}} \bm{\Psi}_{\text{SD}}[n,m] \big),
\end{align}
\vspace*{-2mm}
\hrulefill
\begin{align}\label{eq:MDDMar:yPE} % eq.9
  \bm{y}_{\text{PE}}[n,t,m,m^{\prime}] =&\! \left\{ \!\!\begin{array}{cl}
    \bm{0}, & 1 \leq t \leq 3 \\
    \alpha_{n,m}^{\text{PE}} \widetilde{\bm{F}}_{\text{PE}}^{\rm H}[n,m] \bm{H}_{\text{PE}}^{\rm H}[n,m] \bm{W}_{\text{PE}}[n,m] \bm{r}_{\text{PE}}[n,t,m^{\prime}] + \widetilde{\bm{F}}_{\text{PE}}^{\rm H}[n,m] \bm{n}_{\text{TBS}}[n,m], & t\geq 4 ,
  \end{array} \right. \!\!
\end{align}
\vspace*{-4mm}
\end{figure*}

The received source data at the UAV on subcarrier $m$ is given by (\ref{eq:MDDMar:UAVRxDL}) at the bottom of the page, where \scalebox{0.9}{$\widetilde{\bm{W}}_{\text{SD}}[n,m]\! =\! [\widetilde{\bm{w}}_{\text{SD}}^1[n,m],\cdots ,\widetilde{\pmb{w}}_{\text{SD}}^U[n,m]]\! \in\! \mathbb{C}^{R_{\text{rx}}\times U}$} is the UAV combiner for receiving the source data \scalebox{0.9}{$\bm{x}_{\text{SD}}[n,t,m]\! \in\! \mathbb{C}^U$} from the TBS. Under the TBS power constraint, the precoding matrix \scalebox{0.9}{$\bm{F}_{\text{SD}}[n,m]\! =\! [\pmb{f}_{\text{SD}}^1[n,m],\cdots ,\pmb{f}_{\text{SD}}^U[n,m]]\! \in\! \mathbb{C}^{\bar{R}_{\text{tx}}\times U}$} satisfies \scalebox{0.9}{$\sum_{m\in\mathcal{M}} \alpha_{n,m}^{\text{SD}} \big\|\bm{F}_{\text{SD}}[n,m]\big\|_F^2\! \leq\! P_{\text{TBS}}$}. Due to the strong LoS path and possible scatters around the TBS, the $m$-th subchannel between the TBS and UAV during the $n$-th radio frame follows the Rician distibution, which can be expressed as (\ref{eq:MDDMar:hBU}), where \scalebox{0.9}{$d_{\text{TU}}[n]\! =\! \left\|\bm{c}_{\text{TBS}}\! -\! \bm{c}_{\text{UAV}}[n]\right\|_2$, $\bm{c}_{\text{TBS}}\! =\! \left[x_{\text{TBS}},y_{\text{TBS}},z_{\text{TBS}}\right]$} is the coordinate of TBS, \scalebox{0.9}{$K_{\text{TU}}$} is the Rician factor, and each entry of the small-fading matrix \scalebox{0.9}{$\bm{\Psi}_{\text{SD}}[n,m]$} follows the distribution \scalebox{0.9}{$\mathcal{CN}(0,1)$}. 

As shown in Fig.~\ref{figure-MDDMari-FS}, starting from $t\! =\!4$, the TBS concurrently sends the source data to the UAV and receives the perceived data from the UAV relying on two orthogonal subcarrier sets within the same frequency band, and hence experiences analog-domain SI. Similar to the signal processing at the UAV, we also assume that the TBS is able to provide sufficient analog-domain SIC such that the power of residual analog-domain SI falls into the ADC dynamic range. Then, the residual digital-domain SI can be efficiently canceled by FFT. Consequently, at the TBS receiver, the signal of the $m$-th subcarrier during the $t$-th time slot of the $n$-th radio frame is given in (\ref{eq:MDDMar:yPE}), where \scalebox{0.9}{$\widetilde{\bm{F}}_{\text{PE}}[n,m]\!=\! [\widetilde{\bm{f}}_{\text{PE}}^1[n,m],\cdots ,\widetilde{\bm{f}}_{\text{PE}}^J[n,m]]\! \in\! \mathbb{C}^{\bar{R}_{\text{rx}}\times J}$} is the combining matrix for receiving the perceived signal of targets, \scalebox{0.9}{${\bm{W}}_{\text{PE}}[n,m]\! =\! [{\bm{w}}_{\text{PE}}^1[n,m],\cdots,{\bm{w}}_{\text{PE}}^J[n,m]]\! \in\! \mathbb{C}^{R_{\text{tx}}\times J}$}, \scalebox{0.9}{$\bm{n}_{\text{TBS}}[n,m]\! \sim\! \mathcal{CN}\big(0,N_0\pmb{I}_{\bar{R}_{\text{rx}}}\big)$}, and \scalebox{0.9}{$\bm{H}_{\text{PE}}[n,m]\! \in\! \mathbb{C}^{R_{\text{tx}} \times \bar{R}_{\text{rx}}}$} has a similar form as \scalebox{0.9}{$\bm{H}_{\text{SD}}[n,m]$} in \eqref{eq:MDDMar:hBU}. As the subcarrier used for sensing is orthogonal to that for transmitting perceived signal, the subcarrier $m^{\prime}$ in \scalebox{0.9}{$\bm{r}_{\text{PE}}[n,t,m^{\prime}]$} is different from the subcarrier $m$ in other terms. Furthermore, we assume \scalebox{0.9}{$\left|\mathcal{M}^{\text{SEN}}_n\right|\! =\! \left|\mathcal{M}^{\text{PE}}_n\right|\! =\! M/4$} and there are \scalebox{0.9}{$M/4$} subcarrier pairs, i.e., \scalebox{0.9}{$\mathcal{M}^{\text{pair}}_n\! =\! \left\{\left(m,m^{\prime}\right)| m\! \in\! \mathcal{M}^{\text{PE}}_n, m^{\prime}\! \in\! \mathcal{M}^{\text{SEN}}_n \right\}$}, to sense and forward target information, where any subcarrier in \scalebox{0.9}{$\mathcal{M}^{\text{PE}}_n$} and \scalebox{0.9}{$\mathcal{M}^{\text{SEN}}_n$} can only be paired for once. 

\vspace*{-1mm}
\section{Optimization Problem Design}\label{S3}
\vspace*{-1mm}
By employing the MDD, the subcarriers used for DL communication, sensing and two-way fronthaul transmissions are mutually orthogonal, and the design of their corresponding digital beamforming vectors becomes independent. In order to focus on the joint optimization of UAV trajectory and MDD-enabled resource allocation to maximize the end-to-end communication rate of the proposed network while ensuring the sensing requirements, we adopt the conventional but highly efficient beamforming strategies to reduce the complexity of system optimization.

For two-way fronthaul links, due to the existence of non-negligible non-LoS channels, the ranks of \scalebox{0.9}{$\bm{H}_{\text{SD}}[n,m]$} and \scalebox{0.9}{$\bm{H}_{\text{PE}}[n,m]$} are much larger than the numbers of users and targets, respectively. Performing the singular value decomposition on \scalebox{0.9}{$\bm{H}_{\text{SD}}[n,m]$}~and \scalebox{0.9}{$\pmb{H}_{\text{PE}}[n,m]$ }yields the effective channel matrices~\scalebox{0.9}{$\bar{\bm{H}}_{\text{SD}}[n,m]\!$} \scalebox{0.9}{$=\! \bm{U}_{\text{SD}}[n,m]\bm{\Lambda}^{1/2}_{\text{SD}}[n,m]\bm{L}_{\text{SD}}^{\rm H}[n,m]$}~and \scalebox{0.9}{$\bar{\bm{H}}_{\text{PE}}[n,m]\!$} \scalebox{0.9}{$=\! \bm{U}_{\text{PE}}[n,m]\bm{\Lambda}^{1/2}_{\text{PE}}[n,m]\bm{L}_{\text{PE}}^{\rm H}[n,m]$}, 
where \scalebox{0.9}{$\bm{U}_{\text{SD}}[n,m]\! \in\! \mathbb{C}^{\bar{R}_{\text{tx}}\times U}$} (\scalebox{0.9}{$\bm{U}_{\text{PE}}[n,m]\! \in\! \mathbb{C}^{R_{\text{tx}}\times J}$}) and \scalebox{0.9}{$\bm{L}_{\text{SD}}[n,m]\! \in\! \mathbb{C}^{R_{\text{rx}}\times U}$} (\scalebox{0.9}{$\bm{L}_{\text{PE}}[n,m]\! \in\! \mathbb{C}^{\bar{R}_{\text{rx}}\times U}$}) are the reduced unitary sub-matrices corresponding to the singular matrices of \scalebox{0.9}{$\bm{\Lambda}_{\text{SD}}[n,m]\!$} \scalebox{0.9}{$=\! d_{\text{TU}}^{-2}[n]\text{diag}\left(\eta_{\text{SD}}^1[n,m],\cdots,\eta_{\text{SD}}^U[n,m]\right)$} (\scalebox{0.9}{$\bm{\Lambda}_{\text{PE}}[n,m]\! =\! d_{\text{TU}}^{-2}[n]\text{diag}\left(\eta_{\text{PE}}^1[n,m],\cdots ,\eta_{\text{PE}}^J[n,m]\right)$}). In order to avoid inter-stream interference and maximize the capacity of fronthaul links, we set \scalebox{0.9}{$\bm{F}_{\text{SD}}[n,m]\!$} \scalebox{0.9}{$=\! \bm{U}_{\text{SD}}[n,m]\bm{\Sigma}_{\text{SD}}^{1/2}[n,m]$}, \scalebox{0.9}{$\widetilde{\bm{W}}_{\text{SD}}[n,m]\!$} \scalebox{0.9}{$=\! \bm{L}_{\text{SD}}[n,m]$}, \scalebox{0.9}{$\bm{W}_{\text{PE}}[n,m]\!$} \scalebox{0.9}{$=\! \bm{U}_{\text{PE}}[n,m]{\bm{\Sigma}}_{\text{PE}}^{1/2}[n,m]$}, \scalebox{0.9}{$\widetilde{\bm{F}}_{\text{PE}}[n,m]\!$} \scalebox{0.9}{$=\! \bm{L}_{\text{PE}}[n,m]$}, where \scalebox{0.9}{$\bm{\Sigma}_{\text{SD}}[n,m]\!$} \scalebox{0.9}{$=\! \text{diag}\left(p_{\text{SD}}^1[n,m],\cdots ,p_{\text{SD}}^U[n,m]\right)$} and \scalebox{0.9}{$\bm{\Sigma}_{\text{PE}}[n,m]\!$} \scalebox{0.9}{$=\! \text{diag}\left(p_{\text{PE}}^1[n,m],\cdots ,p_{\text{PE}}^J[n,m]\right)$} denote the matrices of power allocation for source data and perceived sensing data transmissions, respectively. 

\begin{remark}\label{Rk2}
Due to the existence of strong LoS path between the TBS and UAV, for each subcarrier channel $m$ used for transmitting perceived information, there is one singular value much larger than the others in \scalebox{0.9}{$\bm{\Sigma}_{\text{PE}}[n,m]$}. In this case, considering the effective detection of all the targets, the largest singular value with respect to \scalebox{0.9}{$\left|\mathcal{M}^{\text{PE}}_n\right|$} subcarrier channels is equally distributed to \scalebox{0.9}{$J$} targets. In other words, the largest singular value is not always placed at the first diagonal position of \scalebox{0.9}{$\bm{\Sigma}_{\text{PE}}[n,m]$}. Instead, the probability of its occurrence at every diagonal position is the same. By contrast, as the UAV adopts decode-and-forward to pass through the DL signal from TBS to users, we mainly concern the total channel capacity of each subcarrier channel \scalebox{0.9}{$\bm{H}_{\text{SD}}[n,m]$}, and therefore the position of the largest singular value inside \scalebox{0.9}{$\bm{\Sigma}_{\text{SD}}[n,m]$} is inconsequential.   
\end{remark}

To strike a balance between low-complexity and satisfactory  performance, we adopt matched-filtering precoding for communication and~sensing~links.~The~communication precoding matrix is derived as \scalebox{0.9}{$\bm{W}_{\text{DL}}[n,m]\!$} \scalebox{0.9}{$=\! [\bm{w}_{\text{DL}}^1[n,m],\cdots ,\bm{w}_{\text{DL}}^U[n,m]]\!$} \scalebox{0.9}{$=\! [\bm{\alpha }_{\mathrm{tx}}( \theta_{1}^{\mathrm{tx}}[n],\phi_{1}^{\mathrm{tx}}[n],m),\cdots ,\bm{\alpha}_{\mathrm{tx}}( \theta_{U}^{\mathrm{tx}}[n],\phi_{U}^{\mathrm{tx}}[n],m)] \bm{\Sigma}_{\text{DL}}^{1/2}$}, where
\scalebox{0.9}{$\bm{\Sigma}_{\text{DL}}\!$} \scalebox{0.9}{$=\! \text{diag}(p_{\text{DL}}^1[n,m],\cdots ,p_{\text{DL}}^U[n,m])$}~is~the~DL transmission power allocation matrix.~The~{sensing} transmit and~receive~{beamformers}~are implemented as \scalebox{0.9}{$\bm{W}_{\text{SEN}}[n,m]\!$} \scalebox{0.9}{$=\! [\bm{\alpha }_{\mathrm{tx}}( \theta_{1}^{\mathrm{tx}}[n],\phi_{1}^{\mathrm{tx}}[n],m),$} \scalebox{0.9}{$\cdots ,\bm{\alpha }_{\mathrm{tx}}( \theta_{J}^{\mathrm{tx}}[n],\phi_{J}^{\mathrm{tx}}[n],m)\pmb{\Sigma}_{\text{SEN}}^{1/2}$} and \scalebox{0.9}{$\widetilde{\bm{W}}_{\text{SEN}}[n,m]\!$} \scalebox{0.9}{$=\! [\bm{\alpha}_{\mathrm{rx}}\left( \theta_{1}^{\mathrm{rx}}[n],\phi_{1}^{\mathrm{rx}}[n],m\right),\cdots ,\bm{\alpha}_{\mathrm{rx}}( \theta_{J}^{\mathrm{rx}}[n],\phi_{J}^{\mathrm{rx}}[n],m)]$}, where \scalebox{0.9}{$\bm{\Sigma}_{\text{SEN}}[n,m]\!$} \scalebox{0.9}{$=\! \text{diag}\left(p_{\text{SEN}}^1[n,m],\cdots ,p_{\text{SEN}}^J[n,m]\right)$} is the power allocation matrix for target sensing. 

Intuitively, the end-to-end rate is related to the performance of both fronthaul and access transmissions. Denote by \scalebox{0.9}{$R_{\text{SD}}[n]$} and \scalebox{0.9}{$R_{\text{DL}}[n]$} the $n$-th radio frame achievable rate of the TBS sending source data to the UAV and the $n$-th radio frame achievable rate of the UAV transmitting DL data to users, respectively. Based on \eqref{eq:MDDMar:yDL} and \eqref{eq:MDDMar:UAVRxDL}, \scalebox{0.9}{$R_{\text{SD}}[n]$} and \scalebox{0.9}{$R_{\text{DL}}[n]$} can be expressed as follows\setcounter{equation}{9} \vspace{-3mm}

\begin{small}
\begin{align}  % eqs.10,11
  &R_{\text{SD}}[n]\! = \frac{N_\text{s}\! -\! 1}{N_\text{s}}\!\! \sum\nolimits_{m\in \mathcal{M}}\!\!\! \log \det\! \big(\! \bm{I}_U\! +\! \frac{\alpha_{n,m}^{\text{SD}} \bm{\Sigma}_{\text{SD}}[n,m] \bm{\Lambda}_{\text{SD}}[n,m]}{N_0}\! \big)\! ,\label{eq:MDDMar:CDL} \\ 
  &R_{\text{DL}}[n] = \sum\nolimits_{u=1}^U\sum\nolimits_{m \in \mathcal{M}}R^u_{\text{DL}}[n,m], \label{eq:MDDMar:DL}
\end{align}
\end{small}%
with \scalebox{0.9}{$R_{\text{DL}}^u[n,m]\! =\! \frac{N_\text{s}\! -\! 2}{N_\text{s}}\log(1\! +\! \alpha_{n,m}^{\text{DL}}\text{SINR}_{\text{DL}}^u[n,m])$} and \vspace{-3mm} 

\begin{small}
\begin{align}\label{eqSINR} % eq.15
  &\text{SINR}_{\text{DL}}^u[n,m] = \nonumber \\
  & \frac{p_{\text{DL}}^u[n,m]\Omega_{\text{DL}}^u[n,m]}{\sum\nolimits_{u^{'}=1,u^{'}\neq u}^U p_{\text{DL}}^{u^{'}}[n,m]\Omega_{\text{DL}}^{u}[n,m] \left|G_{\text{DL}}^{u^{'},u}[n,m]\right|^2 + N_0} ,
\end{align}
\end{small}% 
in which \scalebox{0.9}{$\Omega_{\text{DL}}^u[n,m]\! =\! (G_{\mathrm{UAV}}^{\mathrm{tx}}G_{\mathrm{UE}}^{\mathrm{rx}}\lambda_m ^2\left|\varpi_{u,m}\right|^2)/(\left( 4\pi \right)^2d_u^{2}[n])$} and \scalebox{0.9}{$G_{\text{DL}}^{u^{'},u}[n,m]\! =\! \bm{\alpha}_{\mathrm{tx}}^{\rm H}\big( \theta^{\mathrm{tx}}_u[n],\phi^{\mathrm{tx}}_u[n],m\big) \bm{\alpha}_{\mathrm{tx}}\big( \theta^{\mathrm{tx}}_{u^{'}}[n],\phi^{\mathrm{tx}}_{u^{'}}[n],m\big)$}.

{To evaluate the sensing performance, we adopt the MI as sensing metric\footnote{Note that, there are multiple targets needed to be sensed on the sea surface, and the sensing task for different targets may be distinct. Hence, using MI as the sensing metric makes the formulation of the overall optimization more generic.}, which exhibits the information-theoretic limit on how much environmental information can be exploited, and hence is widely used in the ISAC literature \cite{shi2017power,ouyang2023integrated}.} As seen in \eqref{eq:MDDMar:UAVRxSenE} and \eqref{eq:MDDMar:yPE}, the objective of sensing is to obtain the information of targets included in \scalebox{0.9}{$\bm{H}_{\text{SEN}}[n,m^{\prime}]$} based on the perceived signal \scalebox{0.9}{$\bm{y}_{\text{PE}}[n,t,m,m^{\prime}]$} at the TBS. Therefore, the end-to-end sensing MI at the $n$-th radio frame is given in \eqref{eq:MDDMar:MI} at the bottom of the next page,
\begin{figure*}[!b]
\footnotesize
\vspace*{-5mm}
\hrulefill
\vspace*{-2mm}
\begin{align}\label{eq:MDDMar:MI} % eq.12
  R_{\text{MI}}[n,m,m^{\prime}] =& \frac{N_\text{s}\! -\! 3}{N_\text{s}} \mathcal{I}\left(\bm{y}_{\text{PE}}[n,t,m,m^{\prime}];\bm{H}_{\text{SEN}}[n,m^{\prime}]|\bm{x}_{\text{SEN}}[n,t,m^{\prime}]\right) = \frac{N_\text{s}\! -\! 3}{N_\text{s}} \log\det\big(\bm{I}_J + \alpha_{m^{\prime},n}^{\text{SEN}} \alpha_{n,m}^{\text{PE}} \bm{\Gamma}^{-1}[n,m,m^{\prime}] \nonumber \\
	& \times \bm{\Lambda}_{\text{PE}}^{1/2}[n,m] \bm{\Sigma}_{\text{PE}}^{1/2}[n,m] \bm{G}[n,m^{\prime}] \bm{\Omega}_{\text{SEN}}[n,m^{\prime}] \bm{\Sigma}_{\text{SEN}}[n,m^{\prime}] \bm{G}[n,m^{\prime}] \bm{\Sigma}_{\text{PE}}^{1/2}[n,m] \bm{\Lambda}_{\text{PE}}^{1/2}[n,m]\big) ,
\end{align}
\vspace*{-2mm}
\end{figure*}
where the entries of \scalebox{0.9}{$\bm{G}[n,m^{\prime}]$} are given by \scalebox{0.9}{$\left(\bm{G}[n,m^{\prime}]\right)_{a,b}\!$} \scalebox{0.9}{$=\! \bm{\alpha}_{\mathrm{rx}}^{\rm H}\left( \theta_{a}^{\mathrm{rx}}[n],\phi_{a}^{\mathrm{rx}}[n],m^{\prime}\right) \bm{\alpha}_{\mathrm{rx}}\left( \theta_{b}^{\mathrm{rx}}[n],\phi_{b}^{\mathrm{rx}}[n],m^{\prime}\right)$}, \scalebox{0.9}{$\forall a,b\! \in\! \mathcal{J}$, $\bm{\Omega}_{\text{SEN}}[n,m^{\prime}]$} is a diagonal matrix with the diagonal entries \scalebox{0.9}{$\left(\bm{\Omega}_{\text{SEN}}[n,m^{\prime}]\right)_{a}\! =\! \frac{G_{\mathrm{UAV}}^{\mathrm{tx}}G_{\mathrm{UAV}}^{\mathrm{rx}}\lambda_{m^{\prime}}^2\sigma^{\mathrm{RCS}}_j\left|\varpi_{j,m^{\prime}}\right|^2}{( 4\pi )^3d_{j}^{4}[n]}$}, \scalebox{0.9}{$\forall a \in \mathcal{J}$}, and \vspace{-5mm}

\begin{small}
\begin{align}\label{eqGamma} % eq.13
  & \bm{\Gamma}[n,m,m^{\prime}] = \bm{\Lambda}_{\text{PE}}^{1/2}[n,m] \bm{\Sigma}_{\text{PE}}^{1/2}[n,m] \widetilde{\bm{W}}_{\text{SEN}}^{\rm H}[n,m^{\prime}]\big( \big( \xi_{\text{SIC}} \nonumber \\
  &\hspace*{2mm} \times \text{Tr}\left(\bm{\Sigma}_{\text{SEN}}[n,m^{\prime}]\right) + N_0\big) \bm{I}_{R_{\text{rx}}}\big) \widetilde{\bm{W}}_{\text{SEN}}[n,m^{\prime}] \bm{\Sigma}_{\text{PE}}^{1/2}[n,m] \nonumber \\
  &\hspace*{2mm} \times \bm{\Lambda}_{\text{PE}}^{1/2}[n,m] + N_0 \bm{I}_J .
\end{align}
\end{small}

According to \eqref{eq:MDDMar:CDL} and \eqref{eq:MDDMar:MI}, the optimization problem can be formulated as \vspace{-3mm}

\begin{small}
\begin{subequations}\label{eq:MDDMar:Opt} % eqs.14a-14j
\begin{align}
  & (\text{P}1)\!:\!\!\!\!\!\! \max_{\left\{\alpha_{n,m}^{\text{X}},\mathcal{M}^{\text{pair}}_n\right\}, \left\{\bm{\Sigma}[n,m]\right\}, \left\{\bm{c}_{\text{UAV}}[n]\right\} } \!\!\!\!\!\!\!\!\!\!\!\!\!\! \min \left\{R_{{\text{SD}}}[n], R_{{\text{DL}}}[n]\right\}\! , \! \label{eqOPob} \\
  & \text{s.t.} \, \alpha_{n,m}^{\text{DL}}\! +\! \alpha_{n,m}^{\text{SEN}}\! +\! \alpha_{n,m}^{\text{PE}}\! +\! \alpha_{n,m}^{\text{SD}}\leq 1,  \forall n, \forall m\in\mathcal{M},\! \label{eqOPc1} \\
  & \hspace*{5mm} \alpha_{n,m}^{\text{X}}\! \in\! \{0,1\},  \forall n, \forall m\in\mathcal{M}, \forall \text{X}\! ,\! \label{eqOPc2} \\
  & \hspace*{5mm} \sum\nolimits_{m \in \mathcal{M}} \alpha_{n,m}^{\text{SEN}}=\sum\nolimits_{m \in \mathcal{M}_n} \alpha_{n,m}^{\text{PE}}=\frac{M}{4}, \forall n, \label{eqOPc3} \\
  & \hspace*{5mm} \sum\nolimits_{m \in \mathcal{M}} \alpha_{n,m}^{\text{SD}} + \sum\nolimits_{m \in \mathcal{M}} \alpha_{n,m}^{\text{DL}} = \frac{M}{2}, \ \forall n, \label{eqOPc4} \\
  & \hspace*{5mm} \sum\nolimits_{m \in \mathcal{M}} R_{\text{DL}}^u[n,m] \geq R_{\text{DL}}^{\text{min}}, \ \forall n, \forall u \in \mathcal{U}, \label{eqOPc5} \\ 
  & \hspace*{5mm} \sum\nolimits_{\left(m,m^{\prime}\right)\in \mathcal{M}^{\text{pair}}_n}\!\! R_{\text{MI}}^j[n,m,m^{\prime}] \geq R_{\text{MI}}^{\text{min}}, \ \forall n, \forall j \in \mathcal{J},\! \label{eqOPc6} \\
  & \hspace*{5mm} \sum\nolimits_{m\in\mathcal{M}}\alpha_{n,m}^{\text{DL}} \text{Tr}\left(\bm{\Sigma}_{\text{DL}}[n,m] \right) + \nonumber \\
	& \hspace*{5mm} \sum\nolimits_{\left(m,m^{\prime}\right)\in \mathcal{M}^{\text{pair}}_n} \big( \alpha_{n,m^{\prime}}^{\text{SEN}}\text{Tr}\left(\bm{\Sigma}_{\text{SEN}}[n,m^{\prime}]\right) + \alpha_{n,m}^{\text{PE}} \alpha_{n,m^{\prime}}^{\text{SEN}} \nonumber \\
	& \hspace*{5mm} \times \text{Tr}\left(\bm{\Pi}_{\text{SEN}}[n,m^{\prime}] \bm{\Sigma}_{\text{PE}}[n,m]\right)\big) \leq P_{\text{UAV}}, \forall n,\! \label{eqOPc7} \\
  & \hspace*{5mm} \sum\nolimits_{m\in\mathcal{M}} \alpha_{n,m}^{\text{SD}} \text{Tr}\left(\bm{\Sigma}_{\text{SD}}[n,m]\right)\leq P_{\text{TBS}}, \forall n, \label{eqOPc8} \\
  & \hspace*{5mm} \left\|\bm{c}_{\text{UAV}}[n]-\bm{c}_{\text{UAV}}[n-1]\right\| \leq V_{\text{max}}N_{\text{s}}T_{\text{s}}, \forall n>1 ,  \label{eqOPc9}
\end{align} 
\end{subequations}
\end{small}%
where \scalebox{0.9}{$\bm{\Pi}_{\text{SEN}}[n,m^{\prime}]\! =\! \bm{G}^{\rm H}[n,m^{\prime}] \bm{\Omega}_{\text{SEN}}[n,m^{\prime}] \bm{\Sigma}_{\text{SEN}}[n,m^{\prime}]$ $\bm{G}[n,m^{\prime}]\!$} \scalebox{0.9}{$+\! \widetilde{\bm{W}}_{\text{SEN}}^{\rm H}[n,m^{\prime}] \big(\left(\xi_{\text{SIC}} \text{Tr}\left(\bm{\Sigma}_{\text{SEN}}[n,m^{\prime}]\right)\! +\! N_0\right) \bm{I}_{R_{\text{rx}}}\big)$ $\widetilde{\bm{W}}_{\text{SEN}}[n,m^{\prime}]$}.

The objective function implies that the final end-to-end DL rate is the minimum of the access rate and fronthaul rate. Constraints (\ref{eqOPc1})-(\ref{eqOPc4}) denote the orthogonality and relative sizes among different subcarrier sets. The user's minimum achievable DL rate requirement at each radio frame is given by (\ref{eqOPc5}). The constraint on the required target estimation accuracy in terms of sensing MI per radio frame is given by (\ref{eqOPc6}), where $R_{\text{MI}}^{\text{min}}$ is the minimum threshold reflecting the basic target characteristics \cite{shi2017power}. The maximum transmission power of the UAV and TBS are constrained by (\ref{eqOPc7}) and (\ref{eqOPc8}), respectively. As the velocity of the UAV is lower than $V_{\text{max}}$, the maximum flight distance between two consecutive time slots is constrained in (\ref{eqOPc9}).

\begin{remark}\label{Rk3}
Different from terrestrial scenarios where on-demand UAV is usually deployed nearby and can carry out transmissions immediately after taking off, the departure point of the UAV in the considered maritime scenario is far away from the destination. Due to the extremely high path loss caused by long distance, the performance of DL communication and especially sensing hardly satisfy the QoS constraints given in (\ref{eqOPc5}) and (\ref{eqOPc6}). Therefore, before handling the optimization problem (P1), it is necessary to first find the initial operating position at which the UAV can simultaneously activate the communication and sensing services toward emergency area. When the UAV arrives at this initial operating position, it can start to work for solving the optimization problem (P1). 
\end{remark}

\vspace*{-4mm}
\section{Solutions to Two Sub-Problems}\label{S4}
As discussed in Remark~\ref{Rk3}, dealing with the problem (P1) amounts to consecutively solve the two sub-problems: (i)~Finding a suitable initial operating position \scalebox{0.9}{$\bm{c}_{\text{UAV}}[1]$} at the first radio frame; (ii)~Starting from this initial operating position, optimizing the trajectory and resource allocation to maximize DL rate while guaranteeing sensing QoS. 
\vspace*{-4mm}
\subsection{First Sub-Problem: Finding Initial Operating Position}\label{S4.1}
We assume that the UAV takes off from the coastal base with the coordinate of \scalebox{0.9}{$\bm{c}_{\text{UAV}}[0]$}. For the purpose of emergency rescue, the UAV is expected to provide prompt communication and sensing services after receiving the command from the TBS. To this end, the distance between \scalebox{0.9}{$\bm{c}_{\text{UAV}}[0]$} and \scalebox{0.9}{$\pmb{c}_{\text{UAV}}[1]$} should be as close as possible, but at the same time, the minimum QoSs of both communication and sensing must be satisfied at \scalebox{0.9}{$\bm{c}_{\text{UAV}}[1]$}. Hence, the first sub-problem can be formulated as follows \vspace{-3mm}

\begin{small}
\begin{subequations}\label{eq:MDDMar:sub1} %eqs.16a-16c
\begin{align}
  (\text{P}2): & \min\limits_{\bm{c}_{\text{UAV}}[1]} \left\|\bm{c}_{\text{UAV}}[1]-\bm{c}_{\text{UAV}}[0]\right\|^2 , \label{eq1SPo} \\
  \text{s.t.} & \ R_{{\text{SD}}}[1] \geq {U\!R}_{\text{DL}}^{\text{min}}, \label{eq1SPc1} \\
  & \ (\ref{eqOPc1})-(\ref{eqOPc8}) \text{ for } n=1 .
\end{align} 
\end{subequations}
\end{small}%
The constraint $(\ref{eq1SPc1})$ ensures that the fronthaul link is able to support the basic DL communication QoS. The optimization (P2) is challenging to solve directly owing to the non-convex constraints (\ref{eqOPc5})-(\ref{eqOPc7}) and the binary integer constraints (\ref{eqOPc1})-(\ref{eqOPc4}). 

\subsubsection{Transform Subcarrier-Related Constraints}\label{MDDMar:sec:3-a}

Recall that there are \scalebox{0.9}{$M/4$} pairs of \scalebox{0.9}{$(m,m^{\prime})\! \in\! \mathcal{M}^{\text{pair}}_n$} with \scalebox{0.9}{$m\! \in\! \mathcal{M}^{\text{PE}}_n$} and \scalebox{0.9}{$m^{\prime}\! \in\! \mathcal{M}^{\text{SEN}}_n$}, where \scalebox{0.9}{$m^{\prime}$} and \scalebox{0.9}{$m$} are used by the UAV to transmit probing signals and convey the collected sensing data to the TBS, respectively. As the performance loss of UAV-target link is much larger than that of UAV-TBS link, to find a suitable initial position, intuitively UAV-target links should have the priority to select subcarriers from \scalebox{0.9}{$\mathcal{M}^{\text{S}}$} such that the sensing QoS can be efficiently satisfied. Hence, for the first radio frame \scalebox{0.9}{$n\! =\! 1$}, \scalebox{0.9}{$\alpha^{\text{SEN}}_{1,m^{\prime}}$} can be determined as follows:\footnote{Although the proposed method is suboptimal in terms of performance, its low complexity is more practical for the UAV scenario. Our future work will study the deep reinforcement learning assisted subcarrier allocation among fronthaul and access links to take both performance and complexity into account.} a.i)~Set \scalebox{0.9}{$\alpha^{\text{SEN}}_{1,m^{\prime}}\! =\!0$}, \scalebox{0.9}{$\forall m^{\prime}\! \in\! \mathcal{M}^{\text{S}}$} and \scalebox{0.9}{$k\! =\! 1$}; a.ii)~Find \scalebox{0.9}{$m^{\prime}_k$} by solving \scalebox{0.9}{$m^{\prime}_k\! =\!\arg\max_{m^{\prime}\! \in\! \mathcal{M}^{\text{S}}} \text{Tr}\left(\pmb{\Omega}_{\text{SEN}}[1,m^{\prime}]\right)$}, and update \scalebox{0.9}{$\alpha^{\text{SEN}}_{1,m^{\prime}_k}\! =\! 1$, $\mathcal{M}^{\text{SEN}}_1\! =\! \mathcal{M}^{\text{SEN}}_1\! \cup\! \left\{m^{\prime}_k\right\}$} and \scalebox{0.9}{$k\! =\! k+1$}; a.iii)~Repeat step a.ii) until \scalebox{0.9}{$\left|\mathcal{M}^{\text{SEN}}_1\right|\! =\! M/4$}. Then, we have \scalebox{0.9}{$\alpha^{\text{PE}}_{1,m}\! =\! 1$}, \scalebox{0.9}{$m\! \in\! \mathcal{M}^{\text{PE}}_1$}, where \scalebox{0.9}{$\mathcal{M}^{\text{PE}}_1\! =\! \mathcal{M}^{\text{S}}\! -\! \mathcal{M}^{\text{SEN}}_1$}. To maximize the end-to-end sensing MI under greedy principle, \scalebox{0.9}{$\mathcal{M}^{\text{pair}}_1$} can be determined as follows: b.i)~Set \scalebox{0.9}{$\mathcal{M}^{\text{pair}}_1\! =\! \emptyset$} and \scalebox{0.9}{$k\! =\! 1$}; b.ii)~Find \scalebox{0.9}{$m_k$} by solving \scalebox{0.9}{$m_k\! =\! \arg\max_{m \in\mathcal{M}^{\text{PE}}_1} \text{Tr}\left(\pmb{\Lambda}_{\text{PE}}[1,m]\right)$}, and update \scalebox{0.9}{$\mathcal{M}^{\text{pair}}_1\! =\! \mathcal{M}^{\text{pair}}_1\! \cup\! \left\{\left(m_k, m^{\prime}_k\right)\right\}$} and \scalebox{0.9}{$k\! =\! k\! +\! 1$}; b.iii)~Repeat step b.ii) until \scalebox{0.9}{$k\! =\! M/4$}. So far, \scalebox{0.9}{$\{\alpha_{1,m}^{\text{PE}},\alpha_{1,m}^{\text{SEN}},\mathcal{M}^{\text{pair}}_1\}$} have been determined. Then $(\ref{eq:MDDMar:Opt}\text{b})$ is reformed as \vspace{-2mm}

\begin{small}
\begin{equation}\label{eq:MDDMar:15b-1} % eq.17
  \alpha_{1,m}^{\text{DL}} + \alpha_{1,m}^{\text{SD}}\leq 1, \  \forall m \in \mathcal{M}^{\text{D}} .
\end{equation}
\end{small}%

\begin{figure*}[!t]\setcounter{equation}{19}
\footnotesize
\vspace*{-4mm}
\begin{align}\label{eqSFA} % eq.20
  &\!\! \Upsilon(\bar{\bm{\Sigma}}_{\text{X}}[1,m])\! =\! 1 - e^{-\frac{\text{Tr}(\bar{\bm{\Sigma}}_{\text{X}}[1,m])}{\rho^{(k)}_{\text{X}}}}\overset{(a)}{\leq} 1 - e^{-\frac{\text{Tr}(\bar{\bm{\Sigma}}^{(k)}_{\text{X}}[1,m])}{\rho^{(k)}_{\text{X}}}}\!\! +\! \frac{1}{\rho^{(k)}_{\text{X}}} e^{-\frac{\text{Tr}(\bar{\bm{\Sigma}}^{(k)}_{\text{X}}[1,m])}{\rho^{(k)}_{\text{X}}}}\! \text{Tr}\big(\bar{\bm{\Sigma}}_{\text{X}}[1,m] - \bar{\bm{\Sigma}}^{(k)}_{\text{X}}[1,m]\big) \triangleq \widetilde{\Upsilon}(\bar{\bm{\Sigma}}_{\text{X}}[1,m]) ,\!
\end{align}
\hrulefill
\vspace*{-2mm}
\setcounter{equation}{24}
\begin{align} % eqs.25,26
  {S}_{\text{MI}}^j[1,m,m^{'}] =& \frac{\eta_{\text{PE}}^j[1,m] p_{\text{PE}}^j[1,m]}{Z_{\text{TU}}[1]}\Big(\sum\nolimits_{j^{'}=1}^J p_{\text{SEN}}^{j^{'}}[1,m^{'}] \bigg(\left|\big(\bm{G}[1,m^{'}]\big)_{j,j^{'}}\right|^2 \frac{\left(\bar{\bm{\Omega}}_{\text{SEN}}[1,m^{\prime}] \right)_{j^{'}}}{Z_{j^{'}}^2[1]} + \xi_{\text{SIC}}\bigg) + N_0 \Big) + N_0, \label{eq:MDDMar:AB} \\
  {N}_{\text{MI}}^j[1,m,m^{'}] =& \frac{\eta_{\text{PE}}^j[1,m]p_{\text{PE}}^j[1,m]}{Z_{\text{TU}}[1]} \Big(\sum\nolimits_{j^{'}=1,j^{'}\neq j}^J \! p_{\text{SEN}}^{j^{'}}[1,m^{'}]\left|\big(\bm{G}[1,m^{'}]\big)_{j,j^{'}}\right|^2\frac{\left(\bar{\bm{\Omega}}_{\text{SEN}}[1,m^{\prime}]\right)_{j^{'}}}{Z_{j^{'}}^2[1]}
+\! \sum\nolimits_{j^{'}=1}^J\! p_{\text{SEN}}^{j^{'}}[1,m^{'}]\xi_{\text{SIC}} +\! N_0\! \Big)  + N_0 , \label{eq:MDDMar:ABC}
\end{align}
\vspace*{-2mm}
\hrulefill
\vspace*{-4mm}
\end{figure*}

The residual binary variables \scalebox{0.9}{$\alpha_{1,m}^{\text{DL}}$} and \scalebox{0.9}{$\alpha_{1,m}^{\text{SD}}$} are coupled with \scalebox{0.9}{$\bm{\Sigma}_{\text{DL}}[1,m]$} and \scalebox{0.9}{$\bm{\Sigma}_{\text{SD}}[1,m]$}, respectively, as shown in (\ref{eqOPc5}), (\ref{eqOPc7}), (\ref{eqOPc8}) and (\ref{eq1SPc1}), which makes these constraints non-convex. Basically, there are four possible combinations between the values of DL-related subcarrier indicators and power matrices, which are \scalebox{0.9}{$\big\{(\alpha_{1,m},\bm{\Sigma}[1,m])|\alpha_{1,m}\! \in\! \{0,1\},\bm{\Sigma}[1,m]\! \in\! \{\bm{0},\bar{\bm{A}}[1,m]\}\big\}$}, where \scalebox{0.9}{$\bar{\bm{A}}[1,m]\! \succeq\! \bm{0}$ and $\bar{\bm{A}}[1,m]\! \neq\! \bm{0}$}. However, it can be seen that only \scalebox{0.9}{$(0,\bm{0})$} and \scalebox{0.9}{$(1,\bar{\bm{A}}[1,m])$} can be the feasible solutions for the problem (P2). Take \scalebox{0.9}{$\alpha_{1,m}^{\text{DL}}$} and \scalebox{0.9}{$\bm{\Sigma}_{\text{DL}}[1,m]$} as an example. If \scalebox{0.9}{$\left(\alpha_{1,m}^{\text{DL}},\bm{\Sigma}_{\text{DL}}[1,m]\right)\! =\! \left(0,\bar{\bm{A}}[1,m]\right)$}, the power allocated to the $m$-th subcarrier have no impact on \scalebox{0.9}{$R_{\text{DL}}[1]$}, and these power will be allocated to other subcarriers to maximize \scalebox{0.9}{$R_{\text{DL}}[1]$}, leading to \scalebox{0.9}{$\bar{\bm{A}}[1,m]\! =\! \bm{0}$} at the end of optimization. If \scalebox{0.9}{$\left(\alpha_{1,m}^{\text{DL}},\bm{\Sigma}_{\text{DL}}[1,m]\right)\! =\! \left(1,\bm{0}\right)$}, the $m$-th subcarrier is wasted as no power is allocated to it. Hence, the optimization is prone to assign subcarrier $m$ to fronthaul link for transmitting source data from TBS to UAV, as the end-to-end DL rate is the minimum of access and fronthaul rates. Thus, we can remove the binary variables by introducing the auxiliary matrices:\setcounter{equation}{17}
\vspace*{-5mm}

\begin{small}
\begin{align}\label{eq:MDDMar:sigDLCD} % eq.18
  %\left\{\!\! \begin{array}{ll}
	  \bar{\bm{\Sigma}}_{\text{DL}}[1,m] & = \text{diag}\left(\bar{p}_{\text{DL}}^1[1,m],\cdots ,\bar{p}_{\text{DL}}^U[1,m]\right)  = \alpha_{1,m}^{\text{DL}} \bm{\Sigma}_{\text{DL}}[1,m], \nonumber \\
    \bar{\bm{\Sigma}}_{\text{SD}}[1,m] & = \text{diag}\left(\bar{p}_{\text{SD}}^1[1,m],\cdots ,\bar{p}_{\text{SD}}^U[1,m]\right)  = \alpha_{1,m}^{\text{SD}} \bm{\Sigma}_{\text{SD}}[1,m].
	%\end{array} \right.\!
\end{align}
\end{small}%
Then \eqref{eq:MDDMar:15b-1} is converted into \vspace{-3mm}

\begin{small}
\begin{equation}\label{eq:MDDMar:18-1} % eq.19
  \left\|\text{Tr}(\bar{\bm{\Sigma}}_{\text{DL}}[1,m])\right\|_0\! +\! \left\|\text{Tr}(\bar{\bm{\Sigma}}_{\text{SD}}[1,m])\right\|_0\! \leq\! 1, \forall m\! \in\! \mathcal{M}^{\text{D}} .\!
\end{equation}
\end{small}%
A smooth function is adopted to approximate the two non-convex \scalebox{0.9}{$\mathcal{L}_0$}-norm functions in \eqref{eq:MDDMar:18-1}, which is given in (\ref{eqSFA}) at the top of the page,
where \scalebox{0.9}{$X\! \in\! \left\{{\text{DL}},{\text{SD}}\right\}$}, (a) is derived by using the first-order Taylor expansion, and \scalebox{0.9}{$\rho^{(k)}_{\text{X}}$} is an iterative smoothing parameter.

\subsubsection{Transform Communication-Related Constraints}

Based on \eqref{eq:MDDMar:sigDLCD}, \scalebox{0.9}{$R_{\text{DL}}[1]$} and \scalebox{0.9}{$R_{\text{SD}}[1]$} are reformulated as\setcounter{equation}{20} \vspace{-3mm} 

\begin{small}
\begin{align}\label{eq:MDDMar:RDLCDR} % eq.21
  \left\{\!\!\! \begin{array}{l}
    R_{\text{DL}}[1]\! =\! \sum\nolimits_{u=1}^{U}\! \underbrace{\frac{N_\text{s}\! -\! 2}{N_\text{s}}\! \sum\nolimits_{m\in \mathcal{M}^{\text{D}}}\! \log\! \left(\! 1\! +\! \frac{{S}_{\text{DL}}^u[1,m]}{{N}_{\text{DL}}^u[1,m]}\right)}_{R_{\text{DL}}^u[1]}, \\
    R_{\text{SD}}[1]\! =\! \sum\nolimits_{u=1}^{U} \frac{N_\text{s}\! -\! 1}{N_\text{s}} \sum\nolimits_{m\in \mathcal{M}^{\text{D}}} \log\left(1\! +\! \frac{{S}_{\text{SD}}^u[1,m]}{{N}_{\text{SD}}^u[1,m]}\right) ,
  \end{array}\right.\!
\end{align}
\end{small}%
where 
\scalebox{0.9}{${S}_{\text{DL}}^u[1,m]\!$} \scalebox{0.9}{$=\!\bar{p}_{\text{DL}}^{u}[1,m]{\bar{\Omega}_{\text{DL}}^{u}[1,m]} \left|G_{\text{DL}}^{u,u}[1,m]\right|^2$}, \scalebox{0.9}{ ${S}_{\text{SD}}^u[1,m]\!$} \scalebox{0.9}{$=\! \eta_{\text{SD}}^u[1,m]p^u_{\text{SD}}[1,m]$}, \scalebox{0.9}{${N}_{\text{DL}}^u[1,m]\!$} \scalebox{0.9}{$=\! \sum\nolimits_{u^{'}=1,u^{'}\neq u}^U \bar{p}_{\text{DL}}^{u^{'}}[1,m]{\bar{\Omega}_{\text{DL}}^{u}[1,m]} \big|G_{\text{DL}}^{u^{'},u}[1,m]\big|^2\! +\! N_0Z_u[1]$} and \scalebox{0.9}{${N}_{\text{SD}}^u[1,m]\!=\! N_0Z_{\text{TU}}[1]$}
with \scalebox{0.9}{$Z_u[1]\! =\! d^2_u[1]\!$} \scalebox{0.9}{$=\! (x_{\text{UAV}}[1]\! -\! x_u)^2\! +\! (y_{\text{UAV}}[1]\! -\! y_u)^2\! +\! (z_{\text{UAV}}\! -\! z_u)^2$}, \scalebox{0.9}{$Z_{\text{TU}}[1]\! =\! d^2_{\text{TU}}\!$} \scalebox{0.9}{$=\! (x_{\text{UAV}}[1]\! -\! x_{\text{TBS}})^2\! +\! (y_{\text{UAV}}[1]\! -\! y_{\text{TBS}})^2\! +\! (z_{\text{UAV}}\! -\! z_{\text{TBS}})^2$} and 
\scalebox{0.9}{$\Omega_{\text{DL}}^{u}[1,m]\! =\! {\bar{\Omega}_{\text{DL}}^{u}[1,m]}/{d^2_u[1]}$}. It can be seen that in \eqref{eq:MDDMar:RDLCDR}, \scalebox{0.9}{$R_{\text{DL}}^u[1]$} is a sum-of-functions-of-ratio problem and \scalebox{0.9}{${{S}_{\text{DL}}^u[1,m]}/{{N}_{\text{DL}}^u[1,m]}$} satisfies the form of concave/convex. Hence, we can convexify the constraint (\ref{eqOPc5}) in an iterative manner with the aid of quadratic transform \cite{shen2018fractional}, which can be expressed as \vspace{-3mm}

\begin{small}
\begin{align}\label{eq:MDDMar:RDLQT} % eq.22
  & \frac{N_\text{s}-2}{N_\text{s}} \sum\nolimits_{m\in \mathcal{M}^{\text{D}}} \log\Big( 1+2v_{\text{DL}}^{u(k)}[1,m]\sqrt{{S}_{\text{DL}}^{u(k+1)}[1,m]} - \nonumber \\
  & \hspace*{2mm} \big(v_{\text{DL}}^{u(k)}[1,m]\big)^2 {N}_{\text{DL}}^{u(k+1)}[1,m]\Big) \triangleq \widetilde{R}_{\text{DL}}^u[1] \geq R_{\text{DL}}^{\min}
\end{align}
\end{small}%
with \scalebox{0.9}{$v_{\text{DL}}^{u(k)}[1,m]\! =\! {\sqrt{{S}_{\text{DL}}^{u(k)}[1,m]}}/{{N}_{\text{DL}}^{u(k)}[1,m]}, \forall u$}. \scalebox{0.9}{$\widetilde{R}_{\text{DL}}^u[1]$} is concave when \scalebox{0.9}{$\left\{v_{\text{DL}}^{u}[1,m]\right\}$} is fixed, and the constraint \eqref{eq:MDDMar:RDLQT} can be iteratively approximated as a convex one. Similarly, the constraint (\ref{eq1SPc1}) can be convexified as \vspace{-3mm}

\begin{small}
\begin{align}\label{eq:MDDMar:RCLQT} % eq. 23
  &  UR_{\text{DL}}^{\text{min}}-  \tilde{R}_{\text{SD}}[1]\leq 0 ,
\end{align}
\end{small}%
where \scalebox{0.9}{$\widetilde{R}_{\text{SD}}[1]\!$} \scalebox{0.9}{$\triangleq \sum_{m\in \mathcal{M}^{\text{D}}} \sum_{u=1}^{U} \frac{N_\text{s}-1}{N_\text{s}} \log(1\! +\! 2v_{\text{SD}}^{u(k)}[1,m]$} \scalebox{0.9}{$\sqrt{{S}_{\text{SD}}^{u(k+1)}[1,m]}\! -\! (v_{\text{SD}}^{u(k)}[1,m])^2 {N}_{\text{SD}}^{u(k+1)}[1,m])$} with \scalebox{0.9}{$v_{\text{SD}}^{u(k)}[1,m]\!$} \scalebox{0.9}{$=\! {\sqrt{{S}_{\text{SD}}^{u(k)}[1,m]}}/{{N}_{\text{SD}}^{u(k)}[1,m]}, \forall u$}.

%\vspace*{-2mm}
\subsubsection{Transform Sensing-Related Constraints}
In order to transform the MI-related constraints (\ref{eqOPc6}) and (\ref{eqOPc7}) into convex ones, we first present the MI of $j$-th target over $(m,m^{'})$ pair, which is given by \vspace{-3mm}

\begin{footnotesize}
\begin{align}\label{eq:MDDMar:MI2} % eq.24
  R_{\text{MI}}^j[1,m,m^{'}] = \frac{N_\text{s}-3}{N_\text{s}} \big( \log\big({{S}_{\text{MI}}^j[1,m,m^{'}]}\big)- \log\big({{N}_{\text{MI}}^j[1,m,m^{'}]} \big) \big) ,
\end{align}
\end{footnotesize}%
where \scalebox{0.9}{${S}_{\text{MI}}^j[1,m,m^{'}]$} and \scalebox{0.9}{${N}_{\text{MI}}^j[1,m,m^{'}]$} are given in \eqref{eq:MDDMar:AB} and \eqref{eq:MDDMar:ABC} at the top of the page, with the diagonal entry
 \scalebox{0.9}{$({\bm{\Omega}}_{\text{SEN}}[1,m^{\prime}])_{j^{'}}\!$}  \scalebox{0.9}{$=\! (\bar{\bm{\Omega}}_{\text{SEN}}[1,m^{\prime}])_{j^{'}} / d_{j^{'}}^4[n]$}, 
\scalebox{0.9}{$Z_{j^{'}}[1]\! =\! d_{j^{'}}^2[1]\!$} \scalebox{0.9}{$=\! (x_\text{UAV}[1]\! -\! x_{j^{'}})^2\! +\! (y_\text{UAV}[1]\! -\! y_{j^{'}})^2\! +\! (z_\text{UAV}\! -\! z_{j^{'}})^2$} and \scalebox{0.9}{$Z_{\text{TU}}[1] = d_{\text{TU}}^{2}[1]\!$} \scalebox{0.9}{$=\! (x_{\text{UAV}}[1]\! -\! x_{\text{TBS}})^2\! +\! (y_{\text{UAV}}[1]\! -\! y_{\text{TBS}})^2\!$} \scalebox{0.9}{$+\! (z_{\text{UAV}}\! -\! z_{\text{TBS}})^2$}. With the aid of the first-order Taylor expansion and successive convex approximation (SCA), \scalebox{0.9}{$\log({S}_{\text{MI}}^j[1,m,m^{'}])$} and \scalebox{0.9}{$\log({N}_{\text{MI}}^j[1,m,m^{'}])$} can be iteratively approximated in a linearized way. The detailed derivation is given in Appendix~\ref{MDDMar:app1}. 
Thus the constraint (\ref{eqOPc6}) is reformulated as\setcounter{equation}{26} \vspace{-4mm}

\begin{footnotesize}
\begin{align}\label{eq:MDDMar:RMICon} % eq.27
\sum\limits_{\left(m,m^{\prime}\right)\in \mathcal{M}^{\text{pair}}_1} \sum\nolimits_{j=1}^J \left(\widetilde{{N}}_{\text{MI}}^j[1,m,m^{'}] - \widetilde{{S}}_{\text{MI}}^j[1,m,m^{'}] \right) + &\frac{N_\text{s}R_{\text{MI}}^{\text{min}}}{N_\text{s}-3}\leq 0 ,
\end{align}
\end{footnotesize}%
which is convex. In addition, the MI-related constraint (\ref{eqOPc7}) can be rewritten as \vspace{-4mm}

\begin{small}
\begin{align}\label{eq:MDDMar:15hre} % eq.28
  & \sum\limits_{m\in \mathcal{M}^{\text{D}}} \text{Tr}(\bar{\bm{\Sigma}}_{\text{DL}}[1,m]) + \sum_{\left(m,m^{\prime}\right)\in \mathcal{M}^{\text{pair}}_1} \big(\text{Tr}({\bm{\Sigma}}_{\text{SEN}}[1,m^{'}]) \nonumber \\
  & + \sum\nolimits_{j=1}^J \widetilde{{P}}_{\text{MI}}^j[1,m,m^{'}]\big) - P_{\text{UAV}} \leq 0 ,
\end{align}\end{small}%
where \scalebox{0.9}{$\widetilde{{P}}_{\text{MI}}^j[1,m,m^{'}]$} is the linear approximation of \scalebox{0.9}{${{P}}_{\text{MI}}^j[1,m,m^{'}]$}, which is given by \vspace{-4mm}

\begin{small}
\begin{align}\label{eq:MDDMar:PMI}
  & {{P}}_{\text{MI}}^j[1,m,m^{'}] = p_{\text{PE}}^j[1,m] \big( \sum\nolimits_{j^{'}=1}^J p_{\text{SEN}}^{j^{'}}[1,m^{'}]\nonumber \\
  & \hspace*{1mm}\times\! \big( \big|(\bm{G}[1,m^{'}])_{j,j^{'}}\big|^2\! \frac{\left(\bar{\bm{\Omega}}_{\text{SEN}}[1,m^{\prime}]\right)_{j^{'}}}{Z_{j^{'}}^2[1]}\! +\! \xi_{\text{SIC}}\! \big)\! +\! N_0\! \big)\! .\!
\end{align}\end{small}%

Based on the above transformation, the problem (P2) can be rewritten as \vspace{-4mm}

\begin{small}
\begin{subequations}\label{eq:MDDMar:sub1-1} % eqs.30a-30h
\begin{align}
  & (\text{P}2.1): \min\limits_{\bm{c}_{\text{UAV}}[1]} \left\|\bm{c}_{\text{UAV}}[1]-\bm{c}_{\text{UAV}}[0]\right\|^2, \label{eqP2.1o} \\
  & \text{s.t.} \ \widetilde{\Upsilon}(\bar{\bm{\Sigma}}_{\text{DL}}[1,m])\! +\! \widetilde{\Upsilon}(\bar{\bm{\Sigma}}_{\text{SD}}[1,m])\leq 1, \forall m \in \mathcal{M}^{\text{D}}, \label{eqP2.1c1} \\
  & \hspace*{6mm} \sum\nolimits_{m\in\mathcal{M}^{\text{D}}}\text{Tr}\left(\bar{\pmb{\Sigma}}_{\text{SD}}[1,m]\right)\leq P_{\text{TBS}}, \label{eqP2.1c2} \\ 
  & \hspace*{6mm} \big|p_{\text{PE}}^{j(k)}[1,m]\! -\! p_{\text{PE}}^{j(k-1)}[1,m]\big|\! \leq\! \vartheta^{(k)}\mu_{\text{p}}, \forall j\! \in\! \mathcal{J}, \forall m\! \in\! \mathcal{M}^{\text{PE}}_1 , \label{eqP2.1c3} \\
  & \hspace*{6mm} \big|p_{\text{SEN}}^{j^{'}(k)}[1,m^{'}]\! -\! p_{\text{SEN}}^{j^{'}(k-1)}[1,m^{'}]\big|\! \leq\! \vartheta^{(k)}\mu_{\text{s}}, \forall j^{'}\! \in\! \mathcal{J}, \forall m^{'}\! \in\! \mathcal{M}^{\text{SEN}}_1 , \label{eqP2.1c4} \\
  & \hspace*{6mm} \big|x_{\text{UAV}}^{(k)}[1] - x_{\text{UAV}}^{(k-1)}[1]\big| \leq \vartheta^{(k)}\mu_{\text{x}}, \label{eqP2.1c5} \\
  & \hspace*{6mm} \big|y_{\text{UAV}}^{(k)}[1] - y_{\text{UAV}}^{(k-1)}[1]\big| \leq \vartheta^{(k)}\mu_{\text{y}}, \label{eqP2.1c6} \\
  & \hspace*{6mm} \eqref{eq:MDDMar:RDLQT}, \eqref{eq:MDDMar:RCLQT}, \eqref{eq:MDDMar:RMICon}, \eqref{eq:MDDMar:15hre} , \label{eqP2.1c7}
\end{align} 
\end{subequations}\end{small}%
where \scalebox{0.9}{$\vartheta^{(k)}\mu_{\text{X}}$} with \scalebox{0.9}{$0<\vartheta^{(k)}\leq 1$} and $\text{X} \in \{\text{p},\text{s},\text{x},\text{y}\}$ denotes the specific radius of the trust region. As the linear approximation based on first-order Taylor expansion is used in $\eqref{eq:MDDMar:RMICon}$ and $\eqref{eq:MDDMar:15hre}$, we impose the constraints (\ref{eqP2.1c3})-(\ref{eqP2.1c6}) to guarantee the accuracy of approximation. 

\subsubsection{Algorithm Implementation}
In (P2.1), the involved QoS constraints are carried out in an iterative manner. To implement iterative optimization, it is crucial to find the initial iteration values of power allocation, and the UAV position, i.e., \scalebox{0.9}{$x_{\text{UAV}}^{(0)}[1]$ and $y_{\text{UAV}}^{(0)}[1]$}, within the feasible region, since both communication and sensing performance depend heavily on the lengths of fronthaul and access links. Hence, during the algorithm implementation, we first initialize \scalebox{0.9}{$\{\bar{p}_{\text{DL}}^{u(0)}[1,m],\bar{p}_{\text{SD}}^{u(0)}[1,m]\}$} and \scalebox{0.9}{$\{p_{\text{SEN}}^{j^{'}(0)}[1,m^{'}]\}$} through equal power allocation. Since $p_{\text{PE}}^{j}[1,m]$ is actually the power coefficient rather than the real power allocated to each target, which is highly related to the UAV position as implied from \eqref{eq:MDDMar:PMI}, \scalebox{0.9}{$\{p_{\text{PE}}^{j(0)}[1,m]\}$} cannot be easily obtained by power equalization. Thus we derive them by solving the following optimization problem: \vspace{-4mm}

\begin{small}
\begin{subequations}\label{eq:MDDMar:sub1-2} % eqs.31.a-31c
\begin{align}
  & (\text{P}2.2): \max_{\left\{p_{\text{PE}}^{j(0)}[1,m]\right\}} \chi , \label{eqP2.2o} \\
  & \text{s.t.} \sum_{\left(m,m^{\prime}\right)\in \mathcal{M}^{\text{pair}}_1}\!\!\!\!\!\! \eta_{\text{PE}}^j[1,m] {{P}}_{\text{MI}}^{j(0)}[1,m,m^{'}] \geq \chi, \forall j\in\mathcal{J}\! ,\!  \label{eqP2.2c1} \\
  & \hspace*{6mm} \sum_{m\in \mathcal{M}^{\text{D}}}\!\!\! \text{Tr}\big(\bar{\bm{\Sigma}}_{\text{DL}}^{(0)}[1,m]\big)\! +\!\!\!\!\! \sum_{\left(m,m^{\prime}\right)\in \mathcal{M}^{\text{pair}}_1}\!\!\!\! \big(\text{Tr}\big({\bm{\Sigma}}_{\text{SEN}}^{(0)}[1,m^{'}]\big)  \nonumber \\
  & \hspace*{16mm} + \sum\nolimits_{j=1}^J{{P}}_{\text{MI}}^{j(0)}[1,m,m^{'}]\big) - P_{\text{UAV}} \leq 0 , \label{eqP2.2c2}
\end{align} 
\end{subequations}\end{small}%
where the constraint (\ref{eqP2.2c1}) ensures that the initialization of \scalebox{0.9}{$p_{\text{PE}}^{j(0)}[1,m]$} maximizes the performance of sensing fronthaul link under fairness consideration. 

\begin{remark}\label{Rk4}
Using the fixed searching step, we can find the feasibly initial iteration value of the UAV position that satisfies all the constraints. The problem (P2.1) aims to minimize the distance between the initial operating point and the take-off point, and the optimal solution may be several kilometer long, which may cause the problem that the first-order Taylor expansion based SCA method is susceptible to get stuck in multiple inflection points during the optimization process. To address this issue, we propose an enhanced local approximation (ELA) method using double-layer iteration to handle the optimization, and the detailed implementation is summarized in Algorithm~\ref{MDDMar:al1}. 
\end{remark}

\begin{algorithm}[t!]
\footnotesize
\caption{\small Find Initial UAV Operating Position} 
\label{MDDMar:al1} % Alg.1
\textbf{Initialization:} \\
  Give UAV taking off position $\bm{c}_{\text{UAV}}[0]$ and designated area center $\bm{c}_{\text{DA}}\! =\! ({x}_{\text{DA}},{y}_{\text{DA}},{z}_{\text{UAV}})$, set position step size $\bm{l}_{\text{step}}$, and thresholds $\epsilon^{'}$ and $\epsilon$\;
  Obtain MI-related subcarrier allocation and pairing results $\mathcal{M}^{\text{PE}}_1$, $\mathcal{M}^{\text{SE}}_1$ and $\mathcal{M}^{\text{pair}}_1$\;
  Set iteration value $k=0$, $d=1$\;
  Obtain $\bar{p}_{\text{DL}}^{u(k)}[1,m]$, $\bar{p}_{\text{SD}}^{u(k)}[1,m]$, $\forall u\in\mathcal{U}, m\in \mathcal{M}^{\text{D}}$ and $p_{\text{SEN}}^{j^{'}(k)}[1,m^{'}]$, $\forall j^{'}\in\mathcal{J}, m^{'}\in \mathcal{M}_1^{\text{SEN}}$ by power equalization\;
  \DT{}{
    \Repeat{\em{(P2.1) is solvable}}{
    Compute $\bm{c}_{\text{UAV}}^{(k)}[1]=\bm{c}_{\text{DA}}-d\cdot \bm{l}_{\text{step}}$\;
    Compute $v_{\text{SD}}^{u(k)}[1,m] = \frac{\sqrt{{S}_{\text{SD}}^{u(k)}[1,m]}}{{N}_{\text{SD}}^{u(k)}[1,m]}, v_{\text{DL}}^{u(k)}[1,m] = \frac{\sqrt{{S}_{\text{DL}}^{u(k)}[1,m]}}{{N}_{\text{DL}}^{u(k)}[1,m]}, \forall u \in \mathcal{U}, m \in \mathcal{M}^{\text{D}}$\;
    Obtain $p_{\text{PE}}^{j(k)}[1,m], \forall j\in\mathcal{J}, m\in \mathcal{M}_1^{\text{PE}}$ by solving optimization problem (P2.2)\;
    Leverage all initial iteration values to deal with optimization problem (P2.1)\;
    \If{\em{(P2.1) is not solvable}}{$d=d+1$\;}
    }
  }
  \QT{}{
    \Repeat{\em{$\left|\bm{c}_{\text{UAV}}^{(k)}[1]-\bm{c}_{\text{UAV}}^{(k-1)}[1]\right|\leq\epsilon$}}{
    Set $k = k + 1$, $k^{'}=1$, $\bm{c}_{\text{UAV}}^{(k^{'})}[1]=\bm{c}_{\text{UAV}}^{(k-1)}[1]$\;
   \Repeat{\em{$\left|\bm{c}_{\text{UAV}}^{(k^{'})}[1]-\bm{c}_{\text{UAV}}^{(k^{'}-1)}[1]\right|\leq\epsilon^{'}$}}{
    Implement Step 5 and Steps 9-10, and then solve optimization problem (P2.1)\;
    Set $k^{'} = k^{'} + 1$\; 
		Update $\bar{p}_{\text{DL}}^{u(k^{'})}[1,m]=\bar{p}_{\text{DL}}^{u(k^{'}-1)}[1,m], \bar{p}_{\text{SD}}^{u(k^{'})}[1,m]=\bar{p}_{\text{SD}}^{u(k^{'}-1)}[1,m], \forall m \in \mathcal{M}^{\text{D}}$, and $p_{\text{SEN}}^{j^{'}(k^{'})}[1,m^{'}]=p_{\text{SEN}}^{j^{'}(k^{'}-1)}[1,m^{'}], p_{\text{PE}}^{j(k^{'})}[1,m]=p_{\text{PE}}^{j(k^{'}-1)}[1,m], \forall \left(m,m^{'}\right) \in \mathcal{M}^{\text{pair}}_1$, and $\bm{c}_{\text{UAV}}^{(k^{'})}[1]=\bm{c}_{\text{UAV}}^{(k^{'}-1)}[1]$\; 
    }
    Update $\bm{c}_{\text{UAV}}^{(k-1)}[1]=\bm{c}_{\text{UAV}}^{(k^{'})}[1]$\;
    }
  }
  \KwOut{$\bm{c}_{\text{UAV}}[1]$.}
\end{algorithm}

\vspace*{-3mm}
\subsection{Second Sub-Problem: Optimization From Initial Operating Position Toward Emergency Area}\label{MDDMar:sec:3-b}

Once the UAV arrives at the initial operating position, it starts to simultaneously carry out communication and sensing tasks with the objective of maximizing the end-to-end DL rate subject to the required sensing QoS, which leads to the second sub-problem expressed as \vspace{-4mm}

\begin{small}
\begin{subequations}\label{eq:MDDMar:sub2} % eqs.32a-32e
\begin{align}
  (\text{P}3): & \max_{\left\{\alpha_{n,m},\mathcal{M}^{\text{pair}}_n\right\}, \left\{\bar{\bm{\Sigma}}[n,m], \bm{\Sigma}[n,m]\right\}, \left\{\bm{c}_{\text{UAV}}[n]\right\}} \psi , \label{eqP3o} \\
  \text{s.t.} & R_{\text{DL}}[n] \geq \psi, \forall n > 1, \label{eqP3c1} \\
  & R_{\text{SD}}[n] \geq \psi, \forall n >1, \label{eqP3c2} \\
  & \sum\nolimits_{m \in \mathcal{M}^{\text{S}}} \alpha_{n,m}^{\text{SEN}}=\sum\nolimits_{m \in \mathcal{M}_n^{\text{S}}} \alpha_{n,m}^{\text{PE}}=\frac{M}{4}, \forall n > 1, \label{eqP3c3} \\
  & (\ref{eqOPc9}), \eqref{eq:MDDMar:RMICon}, \eqref{eq:MDDMar:15hre}, (\ref{eqP2.1c1})-(\ref{eqP2.1c6}) . \label{eqP3c4}
\end{align} 
\end{subequations}\end{small}%
(\ref{eqP3c1}) and (\ref{eqP3c2}) can be convexified by linearizing $R_{\text{DL}}[n]$ and $R_{\text{SD}}[n]$, respectively, following the same method as used in $\eqref{eq:MDDMar:RDLQT}$ and $\eqref{eq:MDDMar:RCLQT}$. 

Next, we have to deal with the constraint (\ref{eqP3c3}), since the MI-related subcarrier allocation in (P3) is totally different from that in (P2). Specifically, after the UAV passing the initial operating position, its relative distances to the TBS and targets are continuously changing, and it is hard to say whether the UAV-target link or UAV-TBS link should have priority in subcarrier allocation.
We adopt the following dynamic allocation method for MI-related subcarriers. After the UAV passing the initial operating position, the UAV-target sensing link has the priority to select subcarriers. Following the same method proposed in Section~\ref{MDDMar:sec:3-a}, we obtain \scalebox{0.9}{$\mathcal{M}^{\text{PE}}_2$} and \scalebox{0.9}{$\mathcal{M}^{\text{SEN}}_2$}. After solving (P3), if the sensing performance improves at the second radio frame, i.e., \scalebox{0.9}{$\sum_{\left(m,m^{\prime}\right)\in \mathcal{M}^{\text{pair}}_{2}}R_{\text{MI}}[2,m,m^{'}]\! -\! \sum_{\left(m,m^{\prime}\right)\in \mathcal{M}^{\text{pair}}_{1}}R_{\text{MI}}[1,m,m^{'}]\! \geq\! 0$}, the same principle is applied at the third radio frame to obtain \scalebox{0.9}{$\mathcal{M}^{\text{PE}}_3$ and $\mathcal{M}^{\text{SEN}}_3$}. Otherwise, at the third radio frame, the two subcarrier sets have to be fine-tuned as follows. First, find \scalebox{0.9}{$m^{\prime}_i\! =\!\arg\max_{m^{\prime} \in \mathcal{M}^{\text{SEN}}_3} \text{Tr}\left(\bm{\Lambda}_{\text{PE}}[3,m]\right)$} and \scalebox{0.9}{$m_j\! =\! \arg\min_{m \in \mathcal{M}^{\text{PE}}_3} \text{Tr}\left(\bm{\Lambda}_{\text{PE}}[3,m]\right)$}. Then, implement \scalebox{0.9}{$\mathcal{M}^{\text{SEN}}_3\! =\! \left(\mathcal{M}^{\text{SEN}}_3\! -\! \left\{m^{\prime}_i\right\} \right)\cup \left\{m_j\right\}$} and \scalebox{0.9}{$\mathcal{M}^{\text{PE}}_3\! =\! \left(\mathcal{M}^{\text{PE}}_3\! -\! \left\{m_j\right\} \right)\cup \left\{m^{\prime}_i\right\}$}. In addition, the pairing between the subcarriers within \scalebox{0.9}{$\mathcal{M}^{\text{SEN}}_3$} and \scalebox{0.9}{$\mathcal{M}^{\text{PE}}_3$} are updated. The process is repeated for \scalebox{0.9}{$n\! >\!  2$} until the end of mission period.

\subsection{Convergence Analysis and Computational Complexity}
\subsubsection{Convergence Analysis}
The convergence of the proposed solutions of the first sub-problem (i.e., (P2.1)) and the second sub-problem (i.e., (P3)) is mainly based on the trust-region aided SCA methods, the convergence of which have been proved in \cite{burden2015numerical}. In particular, as for the optimization of the first sub-problem, after implementing step 6 - step 15 of Algorithm \ref{MDDMar:al1}, the feasibly initial iteration value of the UAV position (i.e., \scalebox{0.9}{$\bm{c}_{\text{UAV}}^{(0)}[1]$}) and its distance with the UAV's take-off point \scalebox{0.9}{$\mathcal{D}^{(0)}=\|\bm{c}_{\text{UAV}}^{(0)}[1]-\bm{c}_{\text{UAV}}[0]\|$} are obtained. Then, within the outer and inner loops of ELA process, as shown in step 17 - step 27 of Algorithm \ref{MDDMar:al1} , we have \scalebox{0.9}{$\mathcal{D}^{(k+1)}\leq \mathcal{D}^{(k)}$} and \scalebox{0.9}{$\mathcal{D}^{(k^{'}+1)}\leq \mathcal{D}^{(k^{'})}, \forall k ~\text{and}~ k^{'}$}, which shows that the value of the object function of (P2.1) is non-increasing over iterations. Additionally, the distance between \scalebox{0.9}{$\bm{c}_{\text{UAV}}[1]$ and $\bm{c}_{\text{UAV}}[0]$} is lower-bounded due to the constraints of communication and sensing QoSs. As for the optimization of the second sub-problem, with the aid of subcarrier iteration and the adjustable radius of trust region \scalebox{0.9}{$\vartheta^{(k)}$}, the objective function of (P3) is non-decreasing over iterations, i.e., \scalebox{0.9}{$\psi^{(k+1)}\geq\psi^{(k)}$} for each iteration \scalebox{0.9}{$k$}, and the value of \scalebox{0.9}{$\psi^{(k)}$} is upper-bounded under the constraints of limited transmit power and flying distance. Therefore, the proposed solutions to the first sub-problem and the second sub-problem are convergent, which is further numerically demonstrated in Section \ref{S5.1}. 

\subsubsection{Computational Complexity}
The computational complexity of solving the first and the second sub-problems are both attributed to three parts, i.e., (i) the computation of beamforming matrices; (ii) MI-related subcarrier allocation and pairing; (iii) the implementation of the involved optimization problems. Since we adopt the low-complexity beamforming design and heuristic method of MI-related subcarrier assignment, the computational complexity of two sub-problems mainly depend on part (iii). Then, according to \cite{nguyen2019joint}, the per-iteration complexity of solving the first and the second sub-problems are computed as \scalebox{0.9}{$\mathcal{O}((\frac{M}{2}+\frac{JM}{2}+7)^{2.5}(\frac{M}{2}+\frac{JM}{2}+11))$} and \scalebox{0.9}{$\mathcal{O}((\frac{M}{2}+\frac{JM}{2}+8)^{2.5}((\frac{UM}{2}+\frac{JM}{2}+2)^{2}+\frac{M}{2}+\frac{JM}{2}+8))$}, respectively.

%In more detail, the optimization of (P2) is mainly based on depend on the optimization of (P2) and (P3), respectively. 

\vspace*{-2mm}
\section{Simulation and Result analysis}\label{S5}
%\vspace*{-2mm}
We consider a coastal emergency scenario centered at \scalebox{0.9}{$(10^4,10^4,0)$}\,m with a radius of \scalebox{0.9}{100}\,m, where \scalebox{0.9}{$U\! =\! 8$} ships and \scalebox{0.9}{$J\! =\! 4$} potential targets are uniformly distributed. The TBS employs two \scalebox{0.9}{$(8\times 8)$} UPAs for transmitting and receiving, respectively, and it is situated at the coordinate of \scalebox{0.9}{$(0,0,10)$}\,m. The UAV equips two \scalebox{0.9}{$(6 \times 6)$} UPAs as transceiver. When receiving the command, it takes off from the coordinate of \scalebox{0.9}{$(10,0,200)$}\,m and is headed to the emergency area, while its flying height is kept at \scalebox{0.9}{200}\,m. Once arriving at the initial operating position, the UAV commences the communication and sensing works lasting for \scalebox{0.9}{$T\! =\!500$}\,s. The number of radio frames within the mission period is \scalebox{0.9}{$N_{\text{t}}\! =\! 500$} and the number of time slots within one radio frame is \scalebox{0.9}{$N_{\text{s}}\! =\! 10$}. Other default system parameters are listed in Table~\ref{Table:MDDMar:para}. 

\begin{table}[t]
\scriptsize
\caption{Default simulation system parameters}
\vspace{-2mm}
\centering
\begin{tabular}{l|l}
\hline
Parameters & Values \\ \hline
TBS and UAV power budget (\scalebox{0.7}{$P_{\text{TBS}}, P_{\text{UAV}}$}) & $(34,30)$ dBm \\ \hline
TBS's transmit and receive antenna gains & (30,26) dBi \\ \hline 
UAV's transmit and receive antenna gains & (24,20) dBi \\ \hline 
User's receive antenna gain (\scalebox{0.7}{$G_{\text{UE}}^{\text{rx}}$}) & 2 dBi \\ \hline
Noise power spectrum density (\scalebox{0.7}{$N_0$}) & -107 dBm \\ \hline
UAV maximum speed (\scalebox{0.7}{$V_{\text{max}}$}) & 30 m/s \\ \hline
Rician factor (\scalebox{0.7}{$K_{\text{TU}}$}) & 30 \\ \hline
RCS (\scalebox{0.7}{$\sigma^{\text{RCS}}_j, \ \forall j$}) & $100$ ${\text{m}}^2$ \\ \hline
Central frequency and bandwidth & $5$ GHz, 10 MHz \\ \hline
Number of subcarriers (\scalebox{0.7}{$M$}) & 32 \\ \hline 
Azimuth AoD/AoA (\scalebox{0.7}{$\phi$}) & $\phi \sim \mathcal{U}(-\pi,\pi)$ \\ \hline
Elevation AoD/AoA (\scalebox{0.7}{$\theta$}) & $\theta \sim \mathcal{U}(-\frac{\pi}{2},\frac{\pi}{2})$ \\ \hline
Minimum communication QoS (\scalebox{0.7}{$R_{\text{DL}}^{\text{min}}$}) & 10 bit/s/Hz \\ \hline
Minimum sensing QoS (\scalebox{0.7}{$R_{\text{MI}}^{\text{min}}$}) & 1 bit/s/Hz \\ \hline
\end{tabular}
\label{Table:MDDMar:para} % Tab.I
\vspace*{-4mm}
\end{table}

\begin{figure}[b!]
\vspace*{-4mm}
\begin{center}
\includegraphics[width=0.6\linewidth]{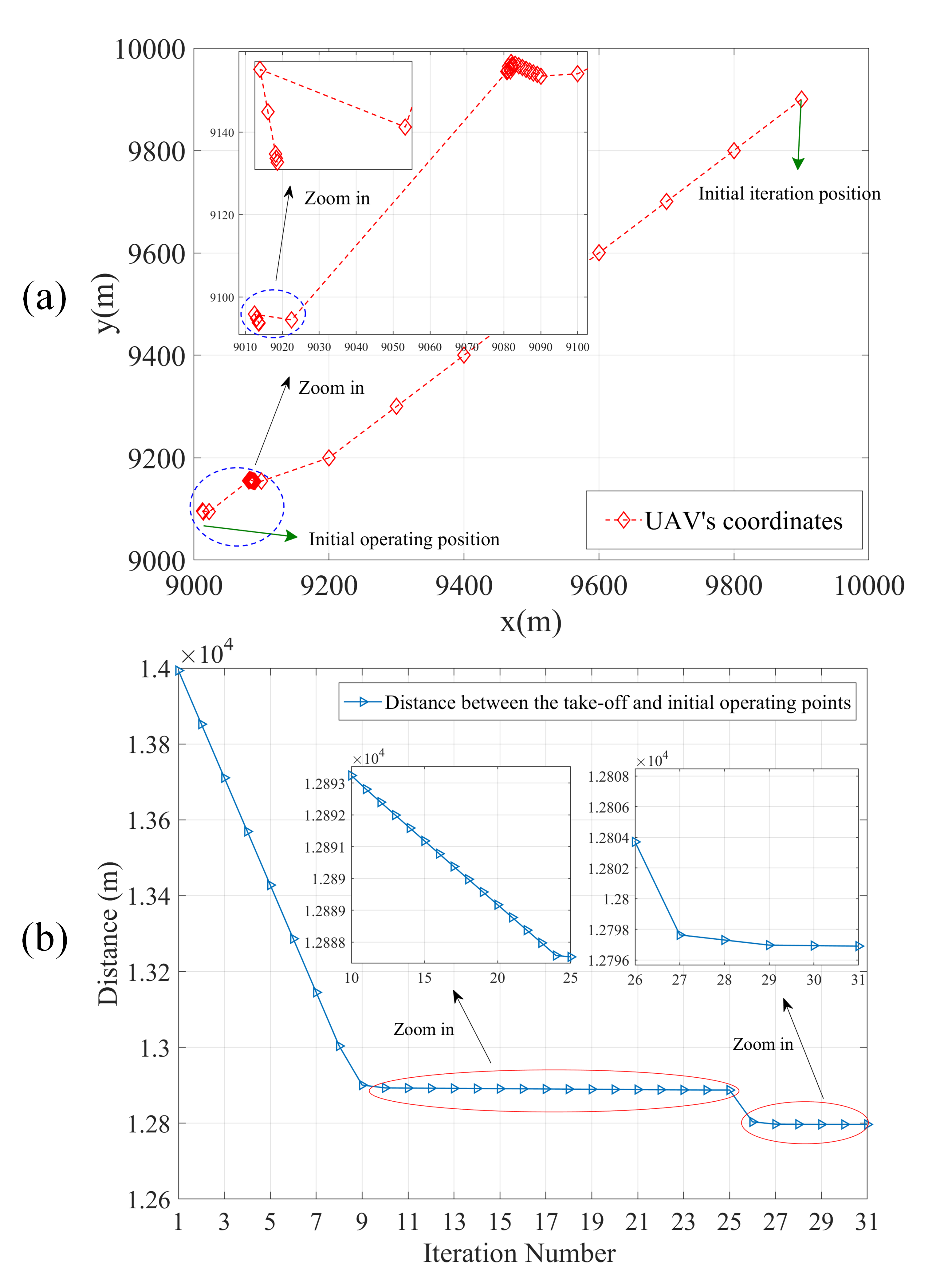}
\end{center}
\vspace{-4mm}
\caption{\small {(a)~ The update process of the UAV's coordinates, and (b)~the distance change between the UAV's take-off and initial operating points, during the implementation of Algorithm 1, under a random network realization.}}
\label{figure-MDDMari-AL1con} % Fig.3
%\vspace*{-1mm}
\end{figure}

\vspace*{-3mm}
{\subsection{Convergence Performance of Two Sub-problems}\label{S5.1}
Given a random network realization, the convergence behavior of the proposed solutions of two sub-problems are depicted as follows. First, as for the first sub-problem, Algorithm~\ref{MDDMar:al1} is applied to find the UAV's initial operating position, and its performance is presented in Fig.~\ref{figure-MDDMari-AL1con}, where Fig.~\ref{figure-MDDMari-AL1con}(a) and Fig.~\ref{figure-MDDMari-AL1con}(b) show the optimization process of UAV's coordinates and objective value over iterations, respectively. In particular, as shown in Fig.~\ref{figure-MDDMari-AL1con}(a), based on steps 6-16 of Algorithm~\ref{MDDMar:al1}, the initial iteration value of the UAV position at $(9900,9900,200)$\,m is determined. Then, by applying the ELA method with a dynamically decreasing factor $\vartheta$, the distance between the UAV's take-off and initial operating points is continuously minimized and converges within 30 iterations. It can be seen from Fig.~\ref{figure-MDDMari-AL1con}(b) that, compared with iteration 1-10, the object value decreases slowly during iteration 10-30. Accordingly, when considering the system overhead of the UAV, the threshold $\vartheta$ can be set to a larger value so as to lower the overall computational complexity. 

%The optimal initial operating position is the closest point from the take-off point that also meets the required CAS QoSs at the first radio frame. %From the second $R_\text{F}$ onward, the distance between the UAV and TBS may change dependent on the real-time CSI. 

\begin{figure}[t]
\vspace*{-5mm}
\begin{center}
\includegraphics[width=0.5\linewidth]{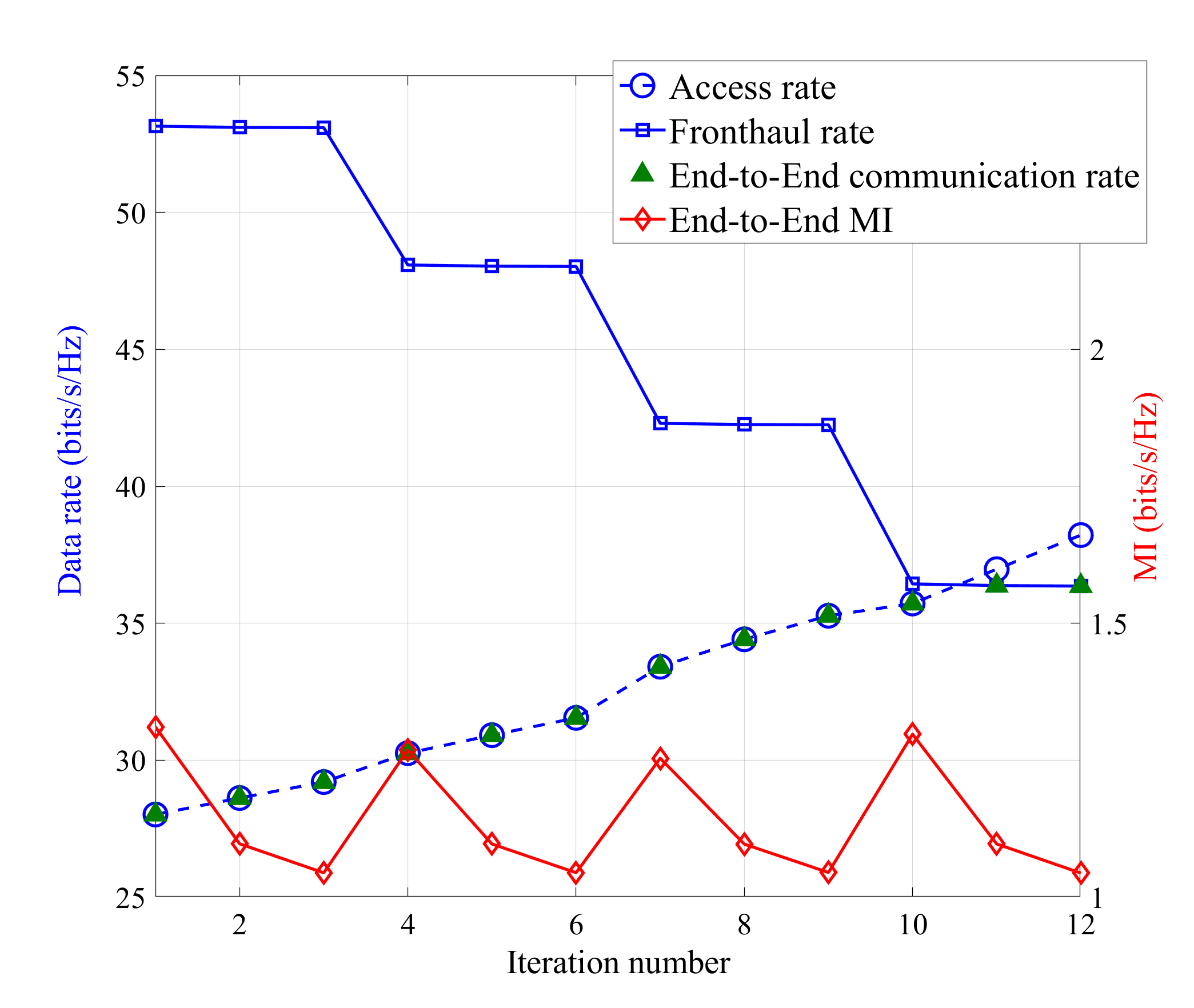}
\end{center}
\vspace{-3mm}
\caption{\small Convergence behavior of the second sub-problem at a random radio frame during mission period.}
\label{figure-MDDMari-conver} % Fig.4
\vspace{-3mm}
\end{figure}
The optimization process of the second sub-problem at one radio frame is plotted in Fig.~\ref{figure-MDDMari-conver}. It can be seen that at the beginning of the iteration process, the access rate is much larger than the fronthaul rate, owing to the fact that the TBS has a higher power budget and the UAV is still far away from the interested area. As the iteration number increases, the UAV moves closer to users and assigns more subcarriers and power to the access links, which enables the end-to-end communication rate to increase until the optimization process converges at the 11-th iteration. Although the MI fluctuates during the iteration process, it is restricted in the feasible region such that the sensing performance is guaranteed.  

\begin{figure}[t]
\vspace*{-1mm}
\begin{center}
\includegraphics[width=0.5\linewidth]{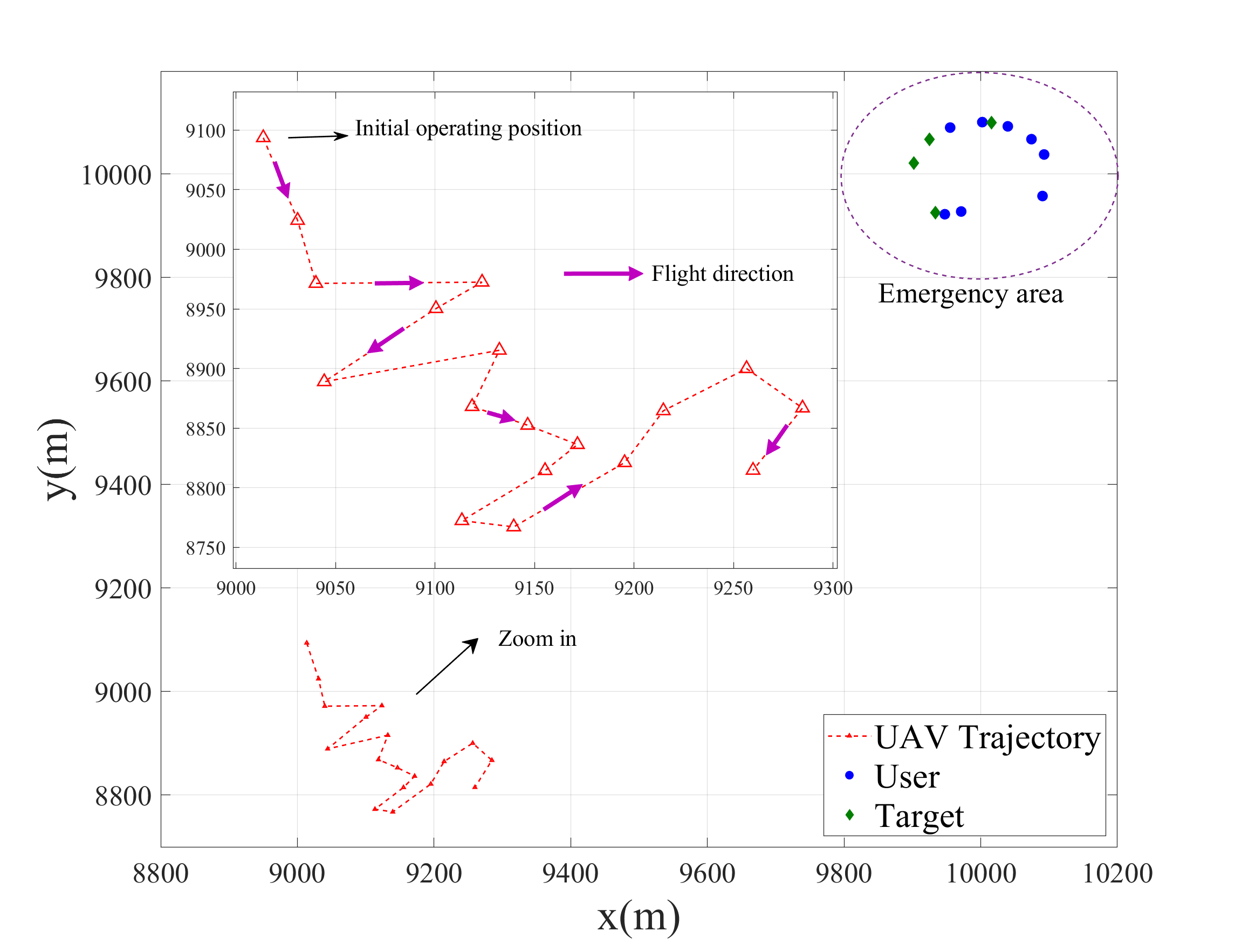}
\end{center}
\vspace{-3mm}
\caption{\small The overall optimal UAV trajectory under a random network realization.}
\label{figure-MDDMari-Trac} % Fig.4
\vspace{-3mm}
\end{figure}

Based on the proposed solutions of two sub-problems, the overall optimal UAV trajectory is plotted in Fig.~\ref{figure-MDDMari-Trac}, where for concise presentation, we uniformly sample 20 points starting from the initial operating position during the whole mission period. It can be seen from Fig.~\ref{figure-MDDMari-Trac} that after the initial operating position, the UAV continuously adjusts the direction and the distance among the TBS and emergency area by considering the effects of fronthaul links, to maximize the end-to-end communication rate while maintaining the required sensing performance.}

\begin{figure}[t]
\vspace*{-4mm}
\begin{center}
\includegraphics[width=0.5\linewidth]{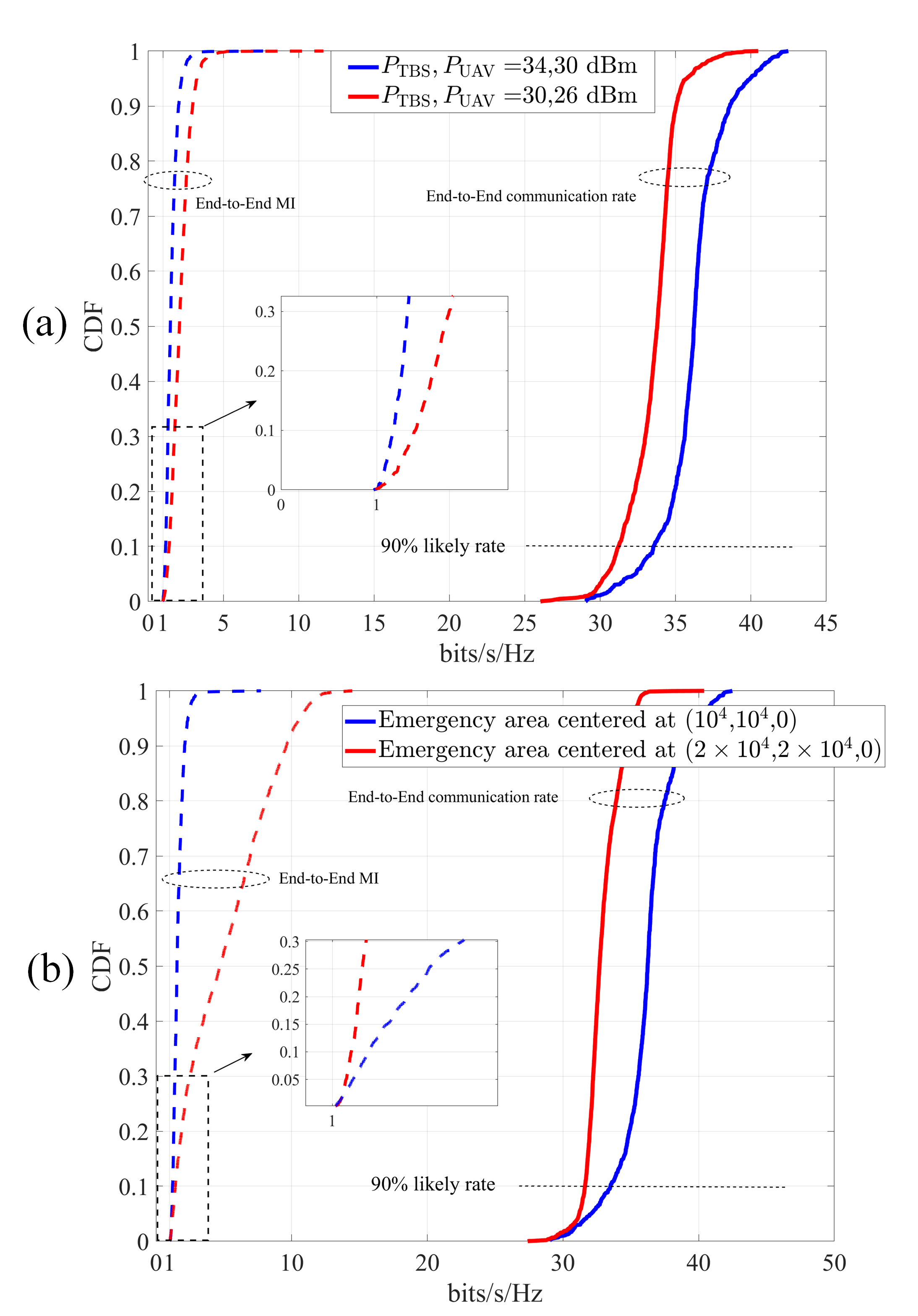}
\end{center}
\vspace{-5mm}
\caption{\small CDF versus end-to-end communication rate and MI in terms of (a) power budgets and (b) emergency area locations.}
\label{figure-MDDMari-power} % Fig.5
\vspace{-3mm}
\end{figure}

\vspace*{-4mm}
\subsection{Performance of Proposed Scheme}\label{S5.2}

We further evaluate the performance of the proposed UAV-ISAC maritime emergency network. To obtain the reliable simulation results, the cumulative distribution function (CDF) performance over thousands of radio frames are evaluated under various network realizations. Fig.~\ref{figure-MDDMari-power}(a) depicts the network performance under two different sets of UAV and TBS power budgets. When the power budget increases, the TBS and UAV can assign more power to transmit source data and DL data via fronthaul and access links, respectively, which lead to higher end-to-end communicate rate. Specifically, an increased of 4\,dBm power at both the UAV and TBS results in an increased 3\,bits/s/Hz at 90\% likely rate. By contrast, increasing the power budget imposes negative effect on the end-to-end MI performance. The reason is because in comparison with sending the sensed information back to the TBS over fronthaul link, the sensing over access links is heavily dependent on the link length, as shown in \eqref{eq:MDDMar:MI}. Therefore, when the power budget is reduced, the UAV is prone to be closer to targets so as to guarantee the required sensing QoS, which leads to an increased end-to-end MI. 

Fig.~\ref{figure-MDDMari-power}(b) studies the impact of emergency area location on the network performance. As expected, the farther the emergency area situates, the smaller the end-to-end communication rate is. When the emergency area is farther away from the coast, the UAV takes the sensing priority over the access link and therefore tends to fly closer to the targets. To meet the sensing QoS constraint, the UAV may allocate more power to sensing subcarriers, resulting in an increase in the end-to-end MI. 

\begin{figure}[b]
\vspace*{-5mm}
\begin{center}
\includegraphics[width=0.6\linewidth]{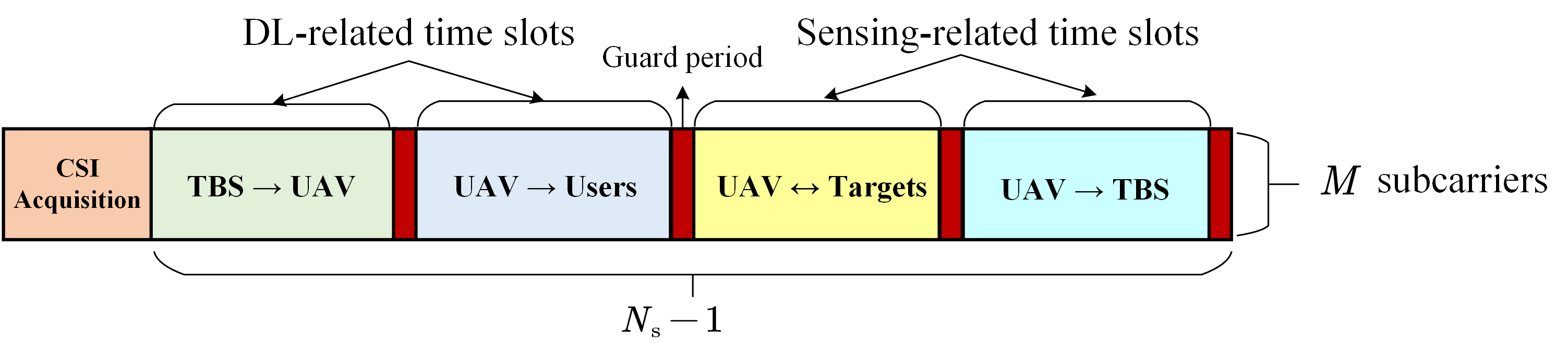}
\end{center}
\vspace*{-5mm}
\caption{\small Radio frame structure of TDMA-UAV-ISAC.}
\label{figure-MDDMari-TDMAFS} % Fig.6
\vspace*{-1mm}
\end{figure}

\begin{figure}[t]
\vspace*{-1mm}
\begin{center}
\includegraphics[width=0.5\linewidth]{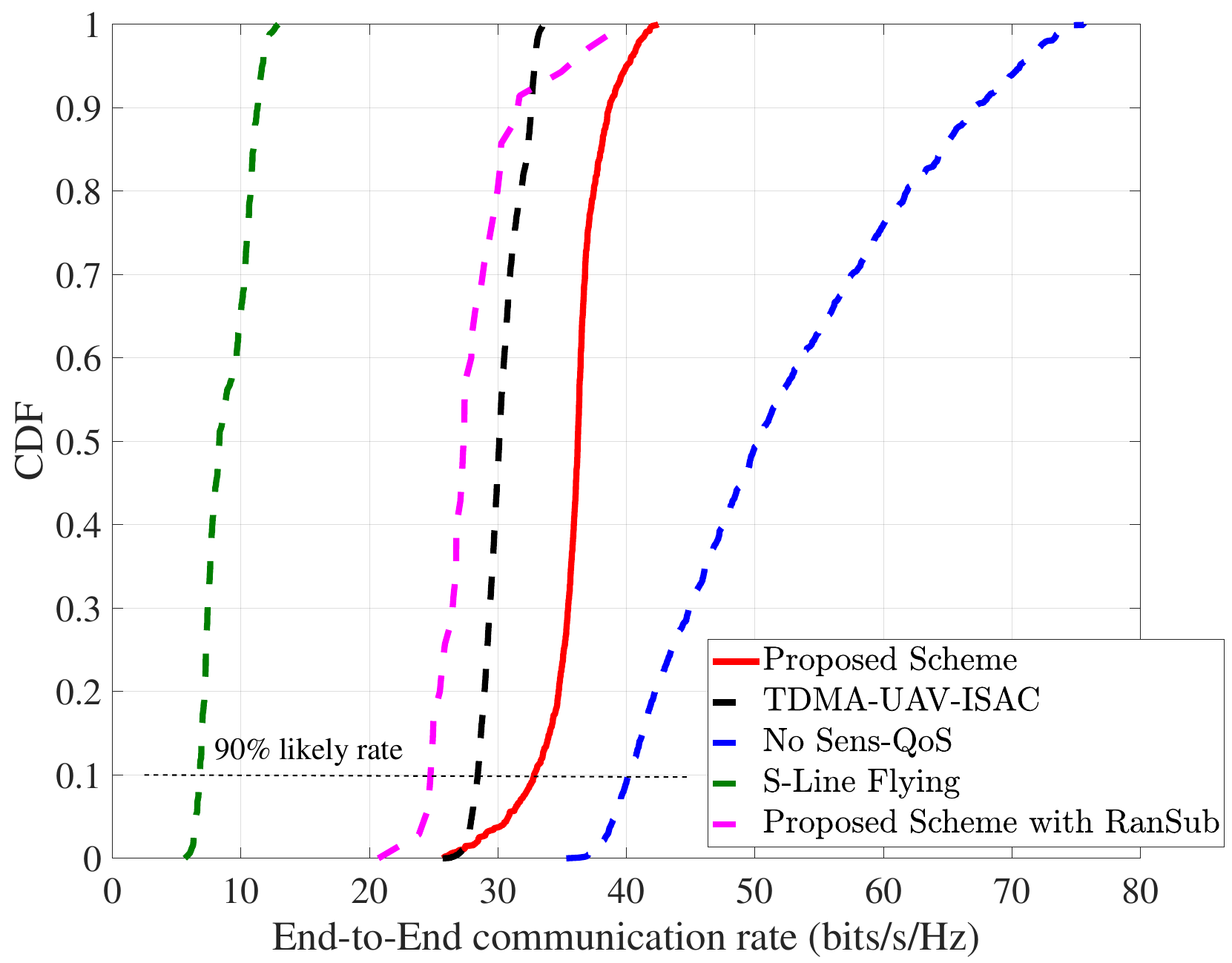}
\end{center}
\vspace*{-4mm}
\caption{\small Network performance comparison for different schemes.}
\label{figure-MDDMari-Compare} % Fig.7
\vspace*{-3mm}
\end{figure}%

Next we compare our UAV-ISAC scheme with the following four benchmarks in maritime emergency networks.
\begin{itemize}
\item {\bf{Proposed scheme with random subcarrier allocation}}\,(Proposed Scheme with RanSub): The subcarrier assignment within \scalebox{0.9}{$\mathcal{M}^{\text{D}}$} and \scalebox{0.9}{$\mathcal{M}^{\text{S}}$} is pre-defined with random selection before the optimization of UAV trajectory and power allocation.
\item {\bf{TDMA-UAV-ISAC}}: The application of TDMA in suppressing the interference between communication and sensing links is widely used in UAV-ISAC related works, such as \cite{liu2023fair,liu2024trajectory,zheng2024dual}. To be compatible with the proposed integrated fronthaul-access networks, the radio frame structure of TDMA-UAV-ISAC is designed as in Fig.~\ref{figure-MDDMari-TDMAFS}. The guard period is essential to avoid the intra-frame interference in TDD-like systems, which accounts for one time slot within each radio frame. 
\item {\bf Straight-Line Flying}\,(S-Line Flying): Once the UAV arrives at the initial operating position, it flies along the line between the initial operating position and the center of emergency area. As the trajectory is fixed, the UAV may not meet the sensing or communication QoS requirement at some positions, in which case the UAV is temporarily out of service. 
\item {\bf No Sensing-QoS Constraint}\,(No Sens-QoS): The UAV carries out DL communication immediately after taking off from the position (10,0,200)\,m, and the sensing QoS constraint is neglected during flying.
\end{itemize}

The performance comparison is presented in Fig.~\ref{figure-MDDMari-Compare}. {No~Sens-QoS} scheme achieves the highest end-to-end communication rate but it may not meet the sensing requirement, since the UAV leverages all the resource to implement DL-related transmission over fronthaul and access links. Our proposed UAV-ISAC achieves 5\,bits/s/Hz more than TDD-UAV-ISAC scheme at 90\% likely rate. It can be attributed to that with the aid of MDD operation and its flexibility in subcarrier allocation, the UAV can simultaneously implement fronthaul and access transmissions at no expense of guard period. The effectiveness of the resource allocation optimization in the proposed scheme is evident as it significantly outperforms Proposed~Scheme~with~RanSub. Without optimizing the UAV's trajectory, S-Line Flying scheme attains the lowest end-to-end communication rate. 
\vspace*{-3mm}
\section{Conclusions}\label{S6}
\vspace*{-1mm}
This paper has investigated the UAV-enabled ISAC technique in maritime emergency networks. Considering that the real-time DL communications and target sensing require the support of robust wireless fronthaul, we have proposed an MDD-based joint fronthaul-access scheme, where the TBS and UAV exchange the source data and perceived information via fronthaul link, and the UAV transmits ISAC signals via access links. To maximize the end-to-end data rate while guaranteeing the required sensing QoS, we have designed an optimization problem to jointly optimize UAV trajectory, subcarrier and power allocation, which is divided into two stages for practical solution, i.e., finding the UAV's initial operating position and optimizing the trajectory and resource allocation in mission period. The SCA and ELA methods have been applied to address the challenging optimization problem. Numerical results have validated that our proposed scheme is capable of balancing the performance between fronthaul and access links such that the communication and sensing services can be properly carried out during UAV's mission. The advantages of our proposed UAV-ISAC scheme over benchmark schemes have also been demonstrated. 

\begin{figure*}[t]\setcounter{equation}{32}
\vspace*{-2mm}
\footnotesize
\begin{align}\label{eq:MDDMar:logSI} % eq.33
  \log\left({{S}_{\text{MI}}^j} \right) \approx & \log\left({{S}_{\text{MI}}^j} \right)|_{\mathcal{Q}_{\text{MI}}^{(k)}} + \frac{1}{\ln^2 {S}_{\text{MI}}^j|_{\mathcal{Q}_{\text{MI}}^{(k)}}} \bigg(\frac{\partial {S}_{\text{MI}}^j}{\partial p_{\text{P}}^{j}}\Big|_{{\mathcal{Q}_{\text{MI}}^{(k)}}} \left(p_{\text{P}}^{j} - p_{\text{P}}^{j(k)}\right) + \sum_{j^{'}} \frac{\partial {S}_{\text{MI}}^j}{\partial p_{\text{S}}^{j^{'}}}\Big|_{{\mathcal{Q}_{\text{MI}}^{(k)}}} \left(p_{\text{S}}^{j^{'}}-p_{\text{S}}^{j^{'}(k)}\right) \nonumber \\
	& + \frac{\partial {S}_{\text{MI}}^j}{\partial x_{\text{U}}}\Big|_{{\mathcal{Q}_{\text{MI}}^{(k)}}} \left(x_{\text{U}} - x_{\text{U}}^{(k)}\right) + \frac{\partial {S}_{\text{MI}}^j}{\partial y_{\text{U}}}\Big|_{{\mathcal{Q}_{\text{MI}}^{(k)}}} \left(y_{\text{U}} - y_{\text{U}}^{(k)}\right) \bigg) \triangleq \widetilde{{S}}_{\text{MI}}^j ,
\end{align}
\vspace*{-1mm}
\hrulefill
\vspace*{-1mm}
\begin{align}\label{eq:MDDMar:SIDe} % eq.34
  & \frac{\partial {S}_{\text{MI}}^j}{\partial p_{\text{P}}^{j}} = \sum\nolimits_{j^{'}=1}^J \frac{\eta_{\text{P}}^j p_{\text{S}}^{j^{'}} \left|(\bm{G})_{j,j^{'}}\right|^2 \left({\bm{\Omega}}_{\text{S}}\right)_{j^{'}}}{Z_{\text{TU}}{Z}_{j^{'}}^2} + \sum\nolimits_{j^{'}=1}^J \frac{\eta_{\text{P}}^j p_{\text{S}}^{j^{'}} \xi_{\text{SIC}}}{Z_{\text{TU}}} + \frac{\eta_{\text{P}}^j N_0}{Z_{\text{TU}}}, ~~
	\frac{\partial {S}_{\text{MI}}^j}{\partial p_{\text{S}}^{j^{'}}} = \frac{\eta_{\text{P}}^j p_{\text{P}}^{j} \left|(\bm{G})_{j,j^{'}}\right|^2 \left({\bm{\Omega}}_{\text{S}}\right)_{j^{'}}}{Z_{\text{TU}}{Z}_{j^{'}}^2} + \frac{\eta_{\text{P}}^j p_{\text{P}}^{j} \xi_{\text{SIC}}}{Z_{\text{TU}}}, \nonumber \\
  & \frac{\partial {S}_{\text{MI}}^j}{\partial x_{\text{U}}} = \sum\nolimits_{j^{'}=1}^J\! -\frac{\eta_{\text{P}}^j p_{\text{P}}^{j} p_{\text{S}}^{j^{'}} \left|(\bm{G})_{j,j^{'}}\right|^2 \left({\bm{\Omega}}_{\text{S}}\right)_{j^{'}}}{Z_{\text{TU}} {Z}_{j^{'}}^2} \bigg(\frac{4\left(x_{\text{U}}\! -\! x_{j^{'}}\right)}{Z_{j^{'}}}\! +\! \frac{2\left(x_{\text{U}}\! -\! x_{\text{T}}\right)}{Z_{\text{TU}}}\bigg) \! -\! \sum\nolimits_{j^{'}=1}^J \frac{2\eta_{\text{P}}^j p_{\text{P}}^{j} p_{\text{S}}^{j^{'}} \xi_{\text{SIC}}\left(x_{\text{U}}\! -\! x_{\text{T}}\right)}{Z_{\text{TU}}^2}\! -\! \frac{\eta_{\text{P}}^j p_{\text{P}}^{j} N_0\left(x_{\text{U}}\! -\! x_{\text{T}}\right)}{Z_{\text{TU}}^2} , \nonumber \\
  & \frac{\partial {S}_{\text{MI}}^j}{\partial y_{\text{U}}} = \sum\nolimits_{j^{'}=1}^J\! -\frac{\eta_{\text{P}}^j p_{\text{P}}^{j} p_{\text{S}}^{j^{'}}\left|(\bm{G})_{j,j^{'}}\right|^2\left({\bm{\Omega}}_{\text{S}}\right)_{j^{'}}}{Z_{\text{TU}}{Z}_{j^{'}}^2}\bigg(\frac{4\left(y_{\text{U}}\! -\! y_{j^{'}}\right)}{Z_{j^{'}}}\! +\! \frac{2\left(y_{\text{U}}\! -\! y_{\text{T}}\right)}{Z_{\text{TU}}}\bigg)\! -\! \sum\nolimits_{j^{'}=1}^J \frac{2\eta_{\text{P}}^j p_{\text{P}}^{j} p_{\text{S}}^{j^{'}} \xi_{\text{SIC}}\left(y_{\text{U}}\! -\! y_{\text{T}}\right)}{Z_{\text{TU}}^2}\! -\! \frac{\eta_{\text{P}}^j p_{\text{P}}^{j} N_0\left(y_{\text{U}}\! -\! y_{\text{T}}\right)}{Z_{\text{TU}}^2} .
\end{align} 
\vspace*{-1mm}
\hrulefill
\vspace*{-1mm}
\begin{align}\label{eq:MDDMar:logNI} % 35
  \log\left({{N}_{\text{MI}}^j} \right) \approx & \log\left({{N}_{\text{MI}}^j} \right)|_{\mathcal{Q}_{\text{MI}}^{(k)}} + \frac{1}{\ln^2 {S}_{\text{NI}}^j|_{\mathcal{Q}_{\text{MI}}^{(k)}}}\bigg(\frac{\partial {N}_{\text{MI}}^j}{\partial p_{\text{P}}^{j}}\Big|_{{\mathcal{Q}_{\text{MI}}^{(k)}}} \left(p_{\text{P}}^{j} - p_{\text{P}}^{j(k)}\right) + \sum_{j^{'}}\frac{\partial {N}_{\text{MI}}^j}{\partial p_{\text{S}}^{j^{'}}}\Big|_{{\mathcal{Q}_{\text{MI}}^{(k)}}}\left(p_{\text{S}}^{j^{'}}-p_{\text{S}}^{j^{'}(k)}\right) \nonumber \\
	& + \frac{\partial {N}_{\text{MI}}^j}{\partial x_{\text{U}}}\Big|_{{\mathcal{Q}_{\text{MI}}^{(k)}}}\left(x_{\text{U}}-x_{\text{U}}^{(k)}\right) + \frac{\partial {N}_{\text{MI}}^j}{\partial y_{\text{U}}}\Big|_{{\mathcal{Q}_{\text{MI}}^{(k)}}}\left(y_{\text{U}}-y_{\text{U}}^{(k)}\right)\bigg) \triangleq \widetilde{{N}}_{\text{MI}}^j ,
\end{align}
\vspace*{-1mm}
\hrulefill
\vspace*{-1mm}
\begin{align}\label{eq:MDDMar:NIDe} % eq.36
  & \frac{\partial {N}_{\text{MI}}^j}{\partial p_{\text{P}}^{j}} = \sum\nolimits_{j^{'}=1,j^{'}\neq j}^J\frac{\eta_{\text{P}}^j p_{\text{S}}^{j^{'}}\left|(\bm{G})_{j,j^{'}}\right|^2 \left({\bm{\Omega}}_{\text{S}}\right)_{j^{'}}}{Z_{\text{TU}}{Z}_{j^{'}}^2} + \sum\nolimits_{j^{'}=1}^J \frac{\eta_{\text{P}}^j p_{\text{S}}^{j^{'}} \xi_{\text{SIC}}}{Z_{\text{TU}}} + \frac{\eta_{\text{P}}^j N_0}{Z_{\text{TU}}}, ~~
  \frac{\partial \bm{N}_{\text{MI}}^j}{\partial p_{\text{S}}^{j^{'}}} =
    \begin{cases}
      \frac{\eta_{\text{P}}^j p_{\text{P}}^{j}\left|(\bm{G})_{j,j^{'}}\right|^2\left({\bm{\Omega}}_{\text{S}}\right)_{j^{'}}}{Z_{\text{TU}}{Z}_{j^{'}}^2} + \frac{\eta_{\text{P}}^j p_{\text{P}}^{j} \xi_{\text{SIC}}}{Z_{\text{TU}}}, j^{'}\neq j , \\
      \frac{\eta_{\text{P}}^j p_{\text{P}}^{j} \xi_{\text{SIC}}}{Z_{\text{TU}}}, j^{'} = j ,
   \end{cases} 
\nonumber \\
  & \frac{\partial {N}_{\text{MI}}^j}{\partial x_{\text{U}}} =\!\!\!\! \sum\nolimits_{j^{'}=1,j^{'}\neq j}^J\!\!\!\! -\frac{\eta_{\text{P}}^j p_{\text{P}}^{j} p_{\text{S}}^{j^{'}}\left|(\bm{G})_{j,j^{'}}\right|^2 \left({\bm{\Omega}}_{\text{S}}\right)_{j^{'}}}{Z_{\text{TU}}{Z}_{j^{'}}^2}\bigg(\frac{4\left(x_{\text{U}}\! -\! x_{j^{'}}\right)}{Z_{j^{'}}}\! +\! \frac{2\left(x_{\text{U}}\! -\! x_{\text{T}}\right)}{Z_{\text{TU}}}\bigg)\! -\! \sum\nolimits_{j^{'}=1}^J\frac{2\eta_{\text{P}}^j p_{\text{P}}^{j} p_{\text{S}}^{j^{'}} \xi_{\text{SIC}}\left(x_{\text{U}}\! -\! x_{\text{T}}\right)}{Z_{\text{TU}}^2}\! -\! \frac{\eta_{\text{P}}^j p_{\text{P}}^{j} N_0\left(x_{\text{U}}\! -\! x_{\text{T}}\right)}{Z_{\text{TU}}^2} , \nonumber \\
  & \frac{\partial {N}_{\text{MI}}^j}{\partial y_{\text{U}}} =\!\!\!\! \sum\nolimits_{j^{'}=1,j^{'}\neq j}^J\!\!\!\! -\frac{\eta_{\text{P}}^j p_{\text{P}}^{j} p_{\text{S}}^{j^{'}}\left|(\bm{G})_{j,j^{'}}\right|^2\left({\bm{\Omega}}_{\text{S}}\right)_{j^{'}}}{Z_{\text{TU}}{Z}_{j^{'}}^2}\bigg(\frac{4\left(y_{\text{U}}\! -\! y_{j^{'}}\right)}{Z_{j^{'}}}\! +\! \frac{2\left(y_{\text{U}}\! -\! y_{\text{T}}\right)}{Z_{\text{TU}}}\bigg)\! -\! \sum\nolimits_{j^{'}=1}^J\frac{2\eta_{\text{P}}^j p_{\text{P}}^{j} p_{\text{S}}^{j^{'}} \xi_{\text{SIC}}\left(y_{\text{U}}\! -\! y_{\text{T}}\right)}{Z_{\text{TU}}^2}\! -\! \frac{\eta_{\text{P}}^j p_{\text{P}}^{j} N_0\left(y_{\text{U}}\! -\! y_{\text{T}}\right)}{Z_{\text{TU}}^2} .\!
\end{align} 
\vspace*{-1mm}
\hrulefill
\vspace*{-4mm}
\end{figure*}

\vspace*{-2mm}
\appendix

\subsection{Linear Approximation of MI-Related Formulation}\label{MDDMar:app1}

For concise expression, the indices of radio frames and subcarriers are temporarily removed, and notations of PE, SEN, UAV and TBS are simplified as P, S, U and T, respectively. Applying SCA, \scalebox{0.9}{$\log({S}_{\text{MI}}^j)$} in \eqref{eq:MDDMar:MI2} can be approximated as (\ref{eq:MDDMar:logSI}) at the top of the next page,
where \scalebox{0.9}{$\mathcal{Q}_{\text{MI}}^{(k)}\! =\! \big(p_{\text{P}}^{j(k)},p_{\text{S}}^{j^{'}(k)},x_{\text{U}}^{(k)},y_{\text{U}}^{(k)}\big)$}, while \scalebox{0.9}{$\frac{\partial {S}_{\text{MI}}^j}{\partial p_{\text{P}}^{j}}$}, \scalebox{0.9}{$\frac{\partial {S}_{\text{MI}}^j}{\partial p_{\text{S}}^{j^{'}}}$}, \scalebox{0.9}{$\frac{\partial {S}_{\text{MI}}^j}{\partial x_{\text{U}}}$} and \scalebox{0.9}{$\frac{\partial {S}_{\text{MI}}^j}{\partial y_{\text{U}}}$} are given in \eqref{eq:MDDMar:SIDe}.
Similarly, \scalebox{0.9}{$\log({N}_{\text{NI}}^j)$} in \eqref{eq:MDDMar:MI2} can be approximated as
\eqref{eq:MDDMar:logNI},
where \scalebox{0.9}{$\mathcal{Q}_{\text{MI}}^{(k)}=\big(p_{\text{P}}^{j(k)},p_{\text{S}}^{j^{'}(k)},x_{\text{U}}^{(k)},y_{\text{U}}^{(k)}\big)$}, while \scalebox{0.9}{$\frac{\partial {N}_{\text{MI}}^j}{\partial p_{\text{P}}^{j}}$}, \scalebox{0.9}{$\frac{\partial {N}_{\text{MI}}^j}{\partial p_{\text{S}}^{j^{'}}}$}, \scalebox{0.9}{$\frac{\partial {N}_{\text{MI}}^j}{\partial x_{\text{U}}}$} and \scalebox{0.9}{$\frac{\partial {N}_{\text{MI}}^j}{\partial y_{\text{U}}}$} are given in \eqref{eq:MDDMar:NIDe}.

%\ifCLASSOPTIONcaptionsoff
%  \newpage
%\fi

\vspace*{-2mm}
\bibliographystyle{IEEEtran}
\bibliography{MDD_Maritime}

\begin{comment}

\end{comment}

%\bibliographystyle{IEEEtran}
%\bibliography{This-paper.bib}

\end{document}